\documentclass[11pt]{article}

\usepackage[latin1]{inputenc}

\usepackage{amstext,amsmath,amssymb,amsfonts,bbm,amsmath}
\usepackage{extarrows}
\usepackage{xcolor} 
\usepackage{hyperref}
\hypersetup{
    colorlinks=false,
    menubordercolor=red,
    linkbordercolor=blue
}
\usepackage{authblk}
\usepackage{caption} 
\usepackage{epsfig} 
\usepackage{color} 
\usepackage{graphicx} 
\usepackage{braket} 
\usepackage{slashed} 
\usepackage{dsfont} 
\usepackage{amsthm}  
\usepackage{multirow}

\usepackage{cite}

\allowdisplaybreaks[4]

\usepackage{psfrag}
\usepackage{tikz} 
\usetikzlibrary{arrows}
\usetikzlibrary{plotmarks}
\usetikzlibrary{external}
\usetikzlibrary{patterns}
\usepackage{pgfplots}
\pgfplotsset{compat=1.9}
\usetikzlibrary{patterns}
\usetikzlibrary{calc}
\usepackage[compat=1.1.0]{tikz-feynman}
\usetikzlibrary{automata,positioning}

\usepackage{float}  
\usepackage{subfig}
\usepackage[lmargin=50pt,rmargin=60pt,tmargin=60pt,bmargin=65pt]{geometry}

\DeclareMathOperator{\Tr}{Tr}

\newcommand{\be}{\begin{equation}}
\newcommand{\ee}{\end{equation}} 

\newcommand{\f}{\frac}
\newcommand{\p}{\partial}

   \let\g=\gamma  
\let\z=\zeta        
    \let\n=\nu           
\let\s=\sigma      
    
\let\G=\Gamma     \let\X=F

\newcommand{\cN}{\mathcal{N}}
\newcommand{\cO}{\mathcal{O}}

\newcommand{\cT}{\mathcal{T}}

\newcommand{\gt}{\tilde{g}}
\newcommand{\rt}{\tilde{r}}

\DeclareMathOperator{\im}{\mathrm{i}}

\newcommand{\mba}{\mathbf{a}}
\newcommand{\mbb}{\mathbf{b}}
\newcommand{\mbc}{\mathbf{c}}
\newcommand{\mbd}{\mathbf{d}}
\newcommand{\mbe}{\mathbf{e}}
\newcommand{\mbf}{\mathbf{f}}
\newcommand{\mbg}{\mathbf{g}}
\newcommand{\mbh}{\mathbf{h}}
\newcommand{\mbm}{\mathbf{m}}
\newcommand{\mbn}{\mathbf{n}}

\allowdisplaybreaks[4]

\numberwithin{equation}{section}

\theoremstyle{remark}

\begin{document}

\title{\bf Long-range multi-scalar models at three loops}

\author[1]{Dario Benedetti}
\author[1,2]{Razvan Gurau}
\author[1]{Sabine Harribey}
\author[1]{Kenta Suzuki}

\affil[1]{\normalsize \it 
 CPHT, CNRS, Ecole Polytechnique, Institut Polytechnique de Paris, Route de Saclay, \authorcr 
\it 91128 PALAISEAU, France
 \authorcr \hfill }

\affil[2]{\normalsize\it 
Perimeter Institute for Theoretical Physics, 31 Caroline St. N, N2L 2Y5, Waterloo, ON,
Canada
 \authorcr
 \bigskip
Emails: dario.benedetti@polytechnique.edu, rgurau@cpht.polytechnique.fr, sabine.harribey@polytechnique.edu,
 kenta.suzuki@polytechnique.edu
 \authorcr \hfill}

\date{}

\maketitle

\hrule\bigskip

\begin{abstract}
We compute the three-loop beta functions of long-range multi-scalar models with general quartic interactions. 
The long-range nature of the models is encoded in a kinetic term with a Laplacian to the power $0<\z<1$, rendering the computation of Feynman diagrams much harder than in the usual short-range case ($\z=1$). As a consequence, previous results stopped at two loops, while seven-loop results are available for short-range models. We push the renormalization group analysis to three loops,
 in an $\epsilon=4\z-d$ expansion at fixed dimension $d<4$, extensively using the Mellin-Barnes representation of Feynman amplitudes in the Schwinger parametrization.
We then specialize the beta functions to various models with different symmetry groups: $O(N)$, $(\mathbb{Z}_2)^N \rtimes S_N$, and $O(N)\times O(M)$. For such models, we compute the fixed points and critical exponents.
\end{abstract}

\hrule\bigskip

\tableofcontents

\section{Introduction}
\label{sec:introduction}

Long-range models have a vast array of applications \cite{Campa:2009rev}, and display several interesting features from a theoretical standpoint. 
In a field theoretic language, they can be characterized as models having in the Lagrangian a kinetic term of the form $\phi (\p^2)^\z\phi$, with $0<\z<1$.\footnote{The restriction to positive $\z$ guarantees standard thermodynamic properties, but models with negative $\z$ are also of phenomenological interest. Models with positive and negative $\z$ are also known as ``weak'' and ``strong'' long-range models, respectively \cite{Mukamel:2009notes}. Notice that in most of the literature a different notation is used for the power of the Laplacian, with $\z\equiv\s/2$.}
Most strikingly, such models admit phase transitions there where they are forbidden in their short-range analogs ($\z=1$), for example in dimension $d=1$, as proved for the long-range Ising model by Dyson \cite{Dyson:1968up}. Moreover, their critical exponents depend on $\z$, thereby defining one-parameter families of universality classes. Tuning this parameter, one can study, at fixed dimension, interesting phenomena such as the transition at some $\z^\star<1$ from a long-range to a short-range universality class \cite{Sak:1973,Blanchard:2012xv,Angelini:2014,Brezin:2014,Defenu:2014,Behan:2017dwr,Behan:2017emf}, or construct rigorous renormalization group results in three dimensions \cite{Brydges:2002wq,Abdesselam:2006qg,Slade:2017,Lohmann:2017}.

The non-integer value of $\z$ brings also some challenges in the analytical treatment of long-range models. For example, the absence of a local energy-momentum tensor complicates the discussion of conformal invariance of the fixed-point theories, an aspect that has only recently been addressed for the long-range Ising model \cite{Paulos:2015jfa}.
Another, very practical, complication arises in the analytic evaluation of Feynman integrals; this will be tackled in the present paper.

Although the interesting features of long-range models have led to their investigation in many different contexts (e.g.\ \cite{Gawedzki:1985jn,Gross:2017vhb,Gubser:2019uyf,Heydeman:2020ijz,Giombi:2019enr,Benedetti:2019eyl,Benedetti:2019ikb,Benedetti:2020yvb, Defenu:2020umv} and other references above or below), they have not been as thoroughly analyzed as their short-range counterparts, partly due to the increased technical  challenges.	
For example, short-range multi-scalar models with quartic interactions have been extensively studied by renormalization group methods both in their general version as well as with various symmetry restrictions (see for example \cite{Pelissetto:2000ek,Kleinert:2001ax,Vicari:2006xr,Osborn:2017ucf,Rychkov:2018vya} and references therein), 
and computations of critical exponents have reached the six-loop approximation \cite{Kompaniets:2017yct},
with the beta function and anomalous dimensions of the $O(N)$ model
being available even at seven loops \cite{Schnetz:2016fhy}. 
Their long-range versions have instead been studied much less, and the renormalization group analysis has been halted at the two-loop computations done in the 1970s \cite{Fisher:1972zz,Yamazaki:1977pt}. Similarly, other methods have also been underdeveloped as compared to the short-range case, with Monte Carlo simulations being mostly limited to the Ising model in one or two dimensions \cite{Glumac:1989,Luijten:1997-thesis,Angelini:2014,uzelac2001critical,tomita2009monte,Luijten:1997,Loscar:2018,rodriguez2009study,Xu:1993}, and with only occasional excursions from other methods, such as the functional renormalization group \cite{Defenu:2014} or the conformal bootstrap \cite{Behan:2018hfx}.

In this paper, we will study the long-range multi-scalar model with quartic interactions. We first compute the beta functions of the general model up to three loops.  
The computation of the Feynman amplitudes of the graphs contributing to the renormalization of the four-point function at three loops is the main result of our paper.
We regularize ultraviolet divergences by setting $\zeta=\frac{d+\epsilon}{4}$ with small $\epsilon$, at fixed dimension $d<4$.\footnote{ A different approach, working directly at finite $\epsilon$ is discussed in \cite{Belim_2003}. The lack of a small-parameter however makes it less rigorous, and prone to the appearance of spurious solutions, as discussed for the short-range case in \cite{Delamotte_2010}.}
 As a renormalization prescription, we use zero-momentum renormalization conditions, and we introduce an infrared regulator $\mu>0$ because  we work with a massless propagator.
All the Feynman amplitudes are computed in the Schwinger parametrization and by exploiting the Mellin-Barnes representation (e.g.\ \cite{Smirnov:2006ry}), with the exception of what we call $I_4$ integral, for which we use the Gegenbauer polynomial technique \cite{Chetyrkin:1980pr} in the momentum space representation.
The general framework and the main results are presented in section \ref{sec:model}, while  some detailed computations are included in the appendices.

We then specialize the beta functions to various symmetry restrictions. 
First, in section \ref{sec:ising model}, we study the long-range Ising model. We give the fixed points and critical exponents in the $\epsilon$ expansion up to order $\epsilon^3$ and compare them with numerical simulations at $d=1$ \cite{Glumac:1989,Luijten:1997-thesis,uzelac2001critical,tomita2009monte} and $d=2$ \cite{Angelini:2014}. 

Second, in section \ref{sec:vector model}, we study the long-range $O(N)$ vector model (the long-range Ising model being the special case $N=1$). We compute the beta functions up to three loops as well as the fixed points and critical exponents and give numerical values for the critical exponents for different dimensions and values of $N$. Our two-loop results agree with \cite{Fisher:1972zz} and \cite{Yamazaki:1977pt}, while the three-loop results are new. We also give the expression for the critical exponents in the large-$N$ expansion. Again our $1/N$ result agrees with \cite{Fisher:1972zz} while the $1/N^2$ contribution is new.

Next, in section \ref{sec:cubic}, we consider the long-range cubic model which is obtained by breaking explicitly the $O(N)$ symmetry with an interaction of the form $\sum_{\mba} \phi_{\mba}^4$. This results in the (hyper-)cubic symmetry group $(\mathbb{Z}_2)^N \rtimes S_N$. We again compute the beta functions, fixed points and critical exponents up to three loops. Our two-loop results agree with \cite{Yamazaki:1978-cubic,Yamazaki:1981-cubic,Chen:2001} while the three-loop results are new. This model admits three-non trivial fixed points: a $O(N)$, or Heisenberg, fixed point with the cubic coupling being zero, an Ising fixed point with the $O(N)$ coupling being zero, and a cubic fixed point when both couplings are non-zero. We then compute the critical value of $N$ at which the Heisenberg and cubic fixed points collapse and exchange stability. 

Finally, in section \ref{sec:bifundamental}, we consider the long-range $O(M)\times O(N)$, or  bifundamental, model. This model has two couplings that are associated to the single-trace and double-trace quartic invariants, in a matrix terminology. We again compute the beta functions and fixed points. However, we do not write explicitly the three loop contributions as they are too lengthy. There are three non-trivial fixed points: a Heisenberg fixed point with the single-trace coupling being zero and two chiral fixed points with both couplings non-zero. We also compute the three critical values of $N$ delimiting four regimes of criticality at fixed $M$, depending on the stability of the Heisenberg and chiral fixed points. We give their expansions up to three loops as well as numerics in three dimensions for $M=2$. For this long-range model we are not aware of any previous results even at two loops.

We conclude in section \ref{sec:concl} with a brief summary and outlook.



\paragraph{Note added.} 
Unfortunately the evaluation of $I_4$ in appendix \ref{app:I_4} contains a mistake that affects $\alpha_{I_4}$ in \eqref{eq:alphas}, and thus all the subsequent numerics. The correct evaluation of the integral and the updated numerics can be found in \cite{Benedetti2024corrig}.

\bigskip
\bigskip

\section{The long-range multi-scalar model}
\label{sec:model}
The long-range multi-scalar model with quartic interactions in dimension $d$ is defined by the action:
\begin{align} \label{eq:action}
		S[\phi]  \, &= \,  \int d^dx \, \bigg[ \frac{1}{2} \phi_\mba(x) ( - \partial^2)^{\zeta}\phi_{\mba}(x) +\frac{1}{2}\, \kappa_{\mba \mbb}\phi_{\mba}(x) \phi_{\mbb}(x) + 
		\frac{1}{4!} \, \lambda_{\mba \mbb \mbc \mbd}
		\phi_{\mba}(x) \phi_{\mbb}(x) \phi_{\mbc}(x) \phi_{\mbd}(x) \bigg] \, ,
\end{align}
where the indices take values from 1 to $\cN$, and a summation over repeated indices is implicit.
The  coupling $\lambda_{\mba\mbb\mbc\mbd}$ and the mass parameter $\kappa_{\mba\mbb}$ are symmetric tensors, thus corresponding in general to $\binom{\cN+3}{4}$ and $\tfrac{\cN(\cN+1)}{2}$ couplings, respectively.
The model is ``long range'' due to the non-integer power of the Laplacian, 
$0< \zeta < 1$.  The short-range model is defined analougously, but with $\z=1$.
From now on the dimension $d$ is fixed to be smaller than (and not necessarily close to) four. 

We treat the mass parameter $\kappa$ as a perturbation, hence the covariance (or propagator) of the free theory is 
$ C_{\mba \mbb}(x,y) =  \delta_{\mba \mbb} \;  C(x-y) $, with:
\be\label{eq:cov} 
\begin{split}
 & C(x-y) = \int\frac{d^dp}{(2\pi)^d}\; e^{ - \im p (x-y) } C(p) = \f{\G\left(\f{d-2\z}{2}\right)}{2^{2\z}\pi^{d/2}\G(\z)} \; \f{1}{|x-y|^{d-2\z}}\,, \\
 & C(p) = \frac{1}{p^{2\zeta}}  = \frac{1}{\Gamma(\zeta)} \int_0^{\infty} da \;a^{\zeta -1 } e^{- a p^2} \,.
\end{split}
\ee 
In the last equality we have introduce also the Schwinger parametrization that we will use to evaluate the Feynman integrals.

The canonical dimension of the field is:
\be
\Delta_{\phi} = \frac{d-2\zeta}{2} \,.
\ee
Therefore, the quartic interaction is irrelevant for $\z<d/4$ leading to mean-field behavior (as rigorously proved in \cite{Aizenman:1988}), while for $\z>d/4$ it is relevant, and a non-trivial IR behavior is expected. The marginal case is $\z=d/4$. We will be interested in the weakly relevant case:
\be \label{eq:zeta-eps}
 \zeta = \frac{d+\epsilon}{4} \,,
\ee
with small $\epsilon$. The ultraviolet dimension of the field is thus fixed to $\Delta_{\phi}=\frac{d-\epsilon}{4}$.

\paragraph{Divergences and regularization.} In order to define an $\epsilon$ expansion, we need to make sense of the theory at $\epsilon=0$ first. However, at $\zeta = d/4$ the model exhibits logarithmic divergences. Let us first consider graphs with $\lambda$ vertices only (we will include the $\kappa$ vertices soon after). 
The two-point graphs are superficially power divergent, the four-point graphs are superficially logarithmically divergent and the higher-point graphs are superficially convergent.

The power divergences can be ignored: one can either use  dimensional regularization or add a two-point power divergent local counterterm to cancel them. Once the local power divergence is subtracted, the two-point graphs are convergent, that is, \emph{there is no wave function renormalization}. This is a feature of the long-range model.

The $\kappa$ vertices represent the $\phi^2$ perturbation with respect to the critical theory. For power counting purposes they can be seen as quartic vertices with two external half edges carrying zero momentum, and thus the previous power counting goes trough with little alteration. The only superficially divergent graphs are:
\begin{itemize}
 \item four-point graphs with only $\lambda$ vertices,
 \item two-point graphs with exactly one $\kappa$ vertex.
\end{itemize}
Both types of graphs are logarithmically divergent, hence we need to regularize both the ultraviolet and the infrared, and choose a renormalization prescription. 

The ultraviolet is naturally regularized by reintroducing $\epsilon>0$.
As a renormalization prescription, we  use the zero momentum BPHZ subtraction scheme \cite{Rivasseau:1991ub}, made explicit in \eqref{eq:bphz} below.
However, since we are working with a massless propagator, an infrared regulator is required. We introduce that by modifying the propagator as: 
\be\label{eq:param}
 C_{\mu}(p) = \frac{1}{(p^2 + \mu^2)^{\zeta}} = \frac{1}{\Gamma(\zeta)} \int_0^{\infty} d a \;a^{\zeta -1 } e^{- a p^2 -a \mu^2} \,,
\ee
for some mass parameter $\mu>0$.
 The reader might wonder why we do not resort to the usual Gell-Mann and Low subtraction at non-zero momentum. The reason is that in such case we were not able to obtain analytic results for the amplitudes of graphs at three loops for $\z<1$.

Before continuing let us comment on the mass term we included in our action. Our regulator $\mu$ is  not a mass: the regulated quadratic part is 
 $\phi(p^2 + \mu^2)^{\zeta}\phi$ while a massive propagator corresponds to
 $\phi(p^{2\zeta} + m^{2\zeta}) \phi $.  $\mu$ and $m$ are identified in the short range case $\zeta=1$, but not in the long range case $\zeta<1$. For the long range model the mass is the coupling of the $\phi^2$ operator, while $\mu$ is relegated to an infrared cutoff\footnote{In the counterterm picture we also tune to criticality by subtracting to zero the mass power divergence, that is we split $m^{2\zeta}$ into a power divergent counterterm and the logarithmic mass coupling $\kappa$. In full detail the quadratic part of the regulated theory is 
$\phi(p^{2} + \mu^2)^{\zeta} \phi + \kappa \phi^2 - m_b^{2\zeta} \phi^2$, with the bare mass  $m_b$ tuned such that $\Braket{\phi(x)\phi(x)}_{\kappa,\mu=0} =0$  
 }. 
 
 The advantage of using the $\mu$ regulated propagator instead of the massive one is threefold. Contrary to the massive propagator ours is analytic at $p^2=0$, it admits a 
 K\"all\'en--Lehmann representation (hence it is Osterwalder Schrader positive):
 \be
  \frac{1}{(p^2+\mu^2)^{\zeta}} \sim \int_{\mu^2}^{\infty} dx \; \frac{ (x-\mu^2)^{-\zeta} }{p^2 + x} \;,
 \ee
 and finally (and most importantly) it allows us to compute the three loop integrals analytically.
 
Once this is settled we have two options. As we are interested in determining the mass critical exponent we can either treat $\kappa$ as a fully fledged coupling of the model (which is what we do below) or we can set it to zero, and look at the anomalous scaling of the renormalized mass operator $[\phi^2]$. We prefer the first option as it treats $\kappa$ and $\lambda$ on the same footing, but we stress that both points of view are equally valid.

\subsection{Two and four-point functions}

We denote $\Gamma^{(2)}_{\mba\mbb}$ and $\Gamma^{(4)}_{\mba \mbb \mbc \mbd}$ the one-particle irreducible two and four-point functions at zero external momentum.\footnote{In the counterterm picture $\Gamma^{(2)}_{\mba\mbb}$ is the two-point function with the local power divergence subtracted.} We compute them up to three loops using the bare expansion in terms of connected amputated one-particle irreducible Feynman diagrams $G$, whose amplitude in Schwinger parametrization reads: 
\be\label{eq:amp_final}
\mathcal{A}(G ) =  \mu^{ (d-4\zeta)(V-1)} \; \mathcal{\hat{A}}(G) \,, \quad
\mathcal{\hat{A}}(G) =
 \frac{1} { 
  \big[ (4\pi)^{d/2} \Gamma(\zeta)^2 \big]^{V-1} }
\int_0^{\infty}
\prod_{e \in G} d a_e
\;\;
\frac{\prod_{e \in G} a_e^{\zeta-1} \; e^{-\sum_{e \in G} a_e}}
{\big(\sum_{\cT \in G  } \prod_{e \notin \cT } a_e\big)^{d/2} } \,,
\ee
where $V$ denote the numbers of vertices of $G$, $e \in G$ runs over the edges of $G$, and $\cT$ denotes the spanning trees in $G$ (e.g.\ \cite{Rivasseau:1991ub}). Note that we used the fact that we only deal with four-point graphs with quartic vertices, as these are sufficient to describe the divergent graphs described above.

\begin{figure}[h!]
	\begin{center}
	\vspace{12pt}
		\includegraphics{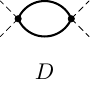} \hspace{20pt} \includegraphics{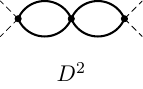} \hspace{20pt} \includegraphics{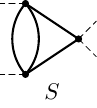}
	\end{center}
	\caption{One and two loops contributions to the bare expansion.}
	\label{fig:1_2_loops}
\end{figure}

\begin{figure}[h!]
	\begin{center}
	\vspace{10pt}
		\includegraphics{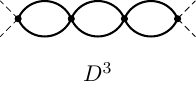} \hspace{20pt} \includegraphics{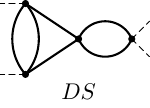} \hspace{20pt} \includegraphics{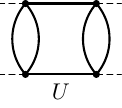} \hspace{20pt} \includegraphics{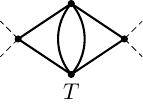} \\[20pt]
		\includegraphics{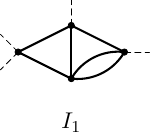} \hspace{20pt} \includegraphics{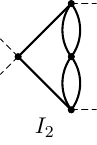} \hspace{20pt} \includegraphics{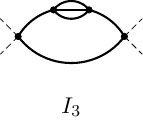} \hspace{20pt} \includegraphics{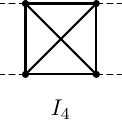}
	\end{center}
	\caption{Three loops contributions.}
	\label{fig:3_loops}
\end{figure}

\paragraph{The four-point function.}
There is only one diagram contributing at one loop, two diagrams at two loops (Fig.~\ref{fig:1_2_loops}), and eight at three loops (Fig.~\ref{fig:3_loops}). We call $D,S,T,U,I_1,I_2,I_3,I_4$ the amplitudes 
$\mathcal{\hat{A}}(G)$ of these diagrams 
(see Fig.~\ref{fig:1_2_loops} and \ref{fig:3_loops} for the detailed notation),\footnote{The choice of letters has no particular meaning, we simply follow the convention of \cite{Benedetti:2019eyl} for the diagrams that had already appeared there.} and we use the fact that the amplitude of a one-vertex reducible diagram (that is, a diagram that disconnects by deleting a vertex) factors into the product of amplitudes of its one-vertex irreducible parts.

One has to be careful to conserve the permutation symmetry of the four-point function in its indices. To this end one should completely symmetrize over the external indices, but due to specific invariances of the diagrams under relabelling, some of the symmetrized terms are trivially equal. Grouping together the terms in classes of explicitly equal terms we get:
\begin{align} \label{eq:bare_4pt}
& \Gamma^{(4)}_{\mba \mbb \mbc \mbd} \, = \, 
\lambda_{\mba \mbb \mbc \mbd}-\frac{1}{2} \big(\lambda_{\mba \mbb \mbe \mbf}\lambda_{\mbe \mbf \mbc \mbd} + 2 \textrm{ terms} \big) \, \mu^{-\epsilon} D \crcr
& 
+\frac{1}{4} \big( \lambda_{\mba \mbb \mbe \mbf}\lambda_{\mbe \mbf \mbg \mbh}\lambda_{\mbg \mbh \mbc \mbd}+ 2 \textrm{ terms} \big) \, \mu^{-2\epsilon} D^2  +\frac{1}{2} (\lambda_{\mba \mbb \mbe \mbf}\lambda_{\mbe \mbg \mbh \mbc}\lambda_{\mbf \mbg \mbh \mbd}+ 5 \textrm{ terms}) \, \mu^{-2\epsilon} S \crcr
& 
- \, \frac{1}{8}  \big(\lambda_{\mba \mbb \mbe \mbf} \lambda_{\mbe \mbf \mbg \mbh} \lambda_{\mbg \mbh \mbm \mbn} \lambda_{\mbm \mbn \mbc \mbd} + 2 \textrm{ terms} \big) \,\mu^{-3\epsilon}  D^3  
- \, \frac{1}{4} \big(\lambda_{\mba \mbb \mbe \mbf} \lambda_{\mbe \mbf \mbg \mbh} \lambda_{\mbg \mbm \mbn \mbc} \lambda_{\mbh \mbm \mbn \mbd} + 5 \textrm{ terms} \big) \, \mu^{-3\epsilon}   SD \crcr
& 
- \, \frac{1}{4} \big(\lambda_{\mba \mbe \mbf \mbg} \lambda_{\mbb \mbe \mbf \mbh} \lambda_{\mbg \mbm \mbn \mbc} \lambda_{\mbh \mbm \mbn \mbd}  + 5 \textrm{ terms} \big) \, \mu^{-3\epsilon} U - \, \frac{1}{4} \big(\lambda_{\mba \mbb \mbe \mbf} \lambda_{\mbe \mbg \mbh \mbm} \lambda_{\mbf \mbg \mbh \mbn} \lambda_{\mbm \mbn \mbc \mbd} + 2 \textrm{ terms} \big) \, \mu^{-3\epsilon} T \crcr
& - \, \frac{1}{2} \big(\lambda_{\mba \mbb \mbe \mbf} \lambda_{\mbe \mbg \mbh \mbm} \lambda_{\mbf \mbg \mbn \mbc} \lambda_{\mbh \mbm \mbn \mbd} + 11 \textrm{ terms} \big) \, \mu^{-3\epsilon} I_1 - 
\,  \frac{1}{4} \big(\lambda_{\mba \mbb \mbe \mbf} \lambda_{\mbe \mbg \mbh \mbc} \lambda_{\mbf \mbm \mbn \mbd} \lambda_{\mbg \mbh \mbm \mbn} + 5 \textrm{ terms} \big) \, \mu^{-3\epsilon}  I_2  \crcr  
&  
- \, \frac{1}{6} \big(\lambda_{\mba \mbb \mbe \mbf}\lambda_{\mbh \mbm \mbn \mbf}\lambda_{\mbh \mbm \mbn \mbg}\lambda_{\mbg \mbe \mbc \mbd} +2 \textrm{ terms} \big)\, \mu^{-3\epsilon} I_3  
-  \,\big(\lambda_{\mba \mbe \mbm \mbh}\lambda_{\mbb \mbe \mbf \mbn}\lambda_{\mbc \mbf \mbm \mbg}\lambda_{\mbd \mbg \mbn \mbh}  \big) \, \mu^{-3\epsilon} I_4  \, .
\end{align}
where the ``$+  \dots \,\rm{terms}$'' notation designates a sum over terms obtained by permuting the external indices 
in non-equivalent ways. For instance, the 3 terms in the first line can be seen as the choice of the index $\mbb$, $\mbc$, or $\mbd$, paired with $\mba$ on the same coupling constant. The $I_1$ diagram is the least symmetric in our lot, only being invariant under exchange of the two external indices attached to the same vertex, thus giving 12 inequivalent terms out of the 24 permutations.

The integrals $D,S,T,U,I_1,I_2,I_3,I_4$ are computed in appendix \ref{app:integrals}.

\paragraph{The two-point function.} The UV-divergent diagrams contributing to
$\Gamma^{(2)}_{\mba\mbb}$ are a subset of those contributing to the four-point function. More specifically, they are the diagrams  in  Fig.~\ref{fig:1_2_loops} and \ref{fig:3_loops} with at least one vertex having two external half edges, hence $U$ and $I_4$ do not contribute. We then substitute one of the vertices  having two external half edges with a $\kappa$ vertex, and we get:
\begin{equation} \label{eq:bare_2pt}
\begin{split}
 \Gamma^{(2)}_{\mbc \mbd} \, = & \, 
\kappa_{ \mbc \mbd}-\frac{1}{2} \big( \kappa_{\mbe \mbf}\lambda_{\mbe \mbf \mbc \mbd} \big) \, \mu^{-\epsilon} D 
+\frac{1}{4} \big( \kappa_{\mbe \mbf} \lambda_{\mbe \mbf \mbg \mbh}\lambda_{\mbg \mbh \mbc \mbd} \big) \, \mu^{-2\epsilon} D^2 
+\frac{1}{2} \big( \kappa_{\mbe \mbf}\lambda_{\mbe \mbg \mbh \mbc}\lambda_{\mbf \mbg \mbh \mbd} \big) \, \mu^{-2\epsilon} S \crcr 
&
- \, \frac{1}{8}  \big( \kappa_{ \mbe \mbf} \lambda_{\mbe \mbf \mbg \mbh} \lambda_{\mbg \mbh \mbm \mbn} \lambda_{\mbm \mbn \mbc \mbd} \big) \,\mu^{-3\epsilon}  D^3  
- \, \frac{1}{4} \big( \kappa_{ \mbe \mbf} \lambda_{\mbe \mbf \mbg \mbh} \lambda_{\mbg \mbm \mbn \mbc} \lambda_{\mbh \mbm \mbn \mbd} \big) \, \mu^{-3\epsilon}   SD \crcr 
& 
- \, \frac{1}{4} \big( \kappa_{ \mbe \mbf} \lambda_{\mbe \mbg \mbh \mbm} \lambda_{\mbf \mbg \mbh \mbn} \lambda_{\mbm \mbn \mbc \mbd} \big) \, \mu^{-3\epsilon} T - \, \frac{1}{2} \big( \kappa_{ \mbe \mbf} \lambda_{\mbe \mbg \mbh \mbm} \lambda_{\mbf \mbg \mbn \mbc} \lambda_{\mbh \mbm \mbn \mbd}  + 1 \textrm{ term} \big) \, \mu^{-3\epsilon} I_1 \crcr 
&
- \,  \frac{1}{4} \big( \kappa_{ \mbe \mbf} \lambda_{\mbe \mbg \mbh \mbc} \lambda_{\mbf \mbm \mbn \mbd} \lambda_{\mbg \mbh \mbm \mbn} \big) \, \mu^{-3\epsilon}  I_2  
- \, \frac{1}{6} \big(\kappa_{ \mbe \mbf}\lambda_{\mbh \mbm \mbn \mbf}\lambda_{\mbh \mbm \mbn \mbg}\lambda_{\mbg \mbe \mbc \mbd}   \big)\, \mu^{-3\epsilon} I_3  \, .
\end{split}
\end{equation}
In general $\Gamma^{(2)}_{\mba\mbb}$ receives contribution also from diagrams with more insertions of $\kappa$ vertices, which are UV-convergent (but IR-divergent); however, we are interested in perturbative fixed-point theories with $\kappa=0$, hence we will ignore such contributions. For the same reason we have not included any $\kappa$ contribution in $\Gamma^{(4)}_{\mba \mbb \mbc \mbd}$. We are only introducing the quadratic operator perturbations as a means to obtain their scaling dimension at the fixed point, for which the linear terms in $\kappa$ suffice.
An equivalent way to rephrase this is to only consider the theory with $\kappa=0$, and in addition study the renormalization of the composite mass quadratic operator $[\phi^2]$ as in \cite{Pelissetto:2000ek,Brezin:1974eb,Brezin:1974-add}.

Observe that the overall combinatorial coefficients in \eqref{eq:bare_2pt} are the same ones as in \eqref{eq:bare_4pt}, but all the symmetries are broken, that is, for each term only one class is selected among the distinct classes involved in the four-point function ($I_1$ is special as it its contribution to $\Gamma^{(2)}_{\mbc\mbd}$ is not a priori symmetric in the indices $\mbc\mbd$, hence one still gets a sum over the two terms). 


\subsection{The beta functions}
\label{sec:beta_func}

In the BPHZ subtraction with IR regulator 
\cite{Rivasseau:1991ub} (which is equivalent to the Wilsonian picture) we identify the dimensionless four-point function at zero external momenta with the running coupling:
\be \label{eq:bphz}
g_{\mba \mbb \mbc \mbd} = \, \mu^{- \epsilon} \, \Gamma^{(4)}_{\mba \mbb \mbc \mbd} \,, \qquad
 r_{\mbc \mbd} = \, \mu^{-(d-2\Delta_{\phi} )} \, \Gamma^{(2)}_{\mbc \mbd} \,.
\ee
The beta functions are the scale derivatives of the running coupling at fixed bare couplings:
\begin{equation}\label{eq:bseries}
\beta^{(4)}_{\mba \mbb \mbc \mbd}=\mu \partial_{\mu} g_{\mba \mbb \mbc \mbd}\, ; \qquad 
 \beta^{(2)}_{\mbc \mbd} = \mu \partial_{\mu} r_{\mbc\mbd} \, .
\end{equation}
We rescale the couplings as\footnote{Notice that $\left(4\pi\right)^{d/2}\Gamma(d/2) =2 (2\pi)^d/{\rm Vol}(S^{d-1})$, where ${\rm Vol}(S^{d-1})= 2\pi^{d/2}/\Gamma(\tfrac{d}{2})$ is the volume of the $(d-1)$-dimensional unit sphere. Our rescaling thus differs by a factor two from the one used by some other authors.}
$g_{\mba \mbb \mbc \mbd}= \left(4\pi\right)^{d/2}\Gamma(d/2) \, \tilde{g}_{\mba \mbb \mbc \mbd} $ and 
$r_{\mba \mbb }= \left(4\pi\right)^{d/2}\Gamma(d/2) \, 
\rt _{\mba \mbb} $, and we denote:  
\begin{align}
 \alpha_{D} \, & =  \, \epsilon (4\pi)^{d/2} \,
\Gamma( \tfrac{d}{2}) \frac{D}{2} \, , &
\alpha_{S} \, &= \, \epsilon (4\pi)^{d}\Gamma( \tfrac{d}{2})^2 \, \frac{(D^2 - 2 S)}{2} \, , \crcr
\alpha_{U} \, & =\, \epsilon (4\pi)^{3d/2}\Gamma(\tfrac{d}{2})^3 \, \frac{(D^3-4DS+3U)}{4} \, , &
\alpha_{T} \,  &=\, \epsilon (4\pi)^{3d/2}\Gamma(\tfrac{d}{2})^3 \, \frac{(3T-2DS)}{4} \, , \;\; \crcr
\alpha_{I_1} \, &   =\, \epsilon (4\pi)^{3d/2}\Gamma(\tfrac{d}{2})^3 \, \frac{(D^3-3DS+3I_1)}{2} \, , &
\alpha_{I_2} \,  & = \, \epsilon (4\pi)^{3d/2}\Gamma(\tfrac{d}{2})^3\frac{(D^3-4DS+3I_2)}{4} \, , \crcr
\alpha_{I_3} \, &  =\, \epsilon (4\pi)^{3d/2}\Gamma(\tfrac{d}{2})^3 \, \frac{I_3}{2} \, , &
\alpha_{I_4} \, & =\, \epsilon  (4\pi)^{3d/2}\Gamma(\tfrac{d}{2})^3 \,  3I_4 \, \,.
\end{align}
Using appendix \ref{app:renseries}, we get:
\begin{align}
\beta^{(4)}_{\mba \mbb \mbc \mbd} &= -\epsilon \gt_{\mba \mbb \mbc \mbd} + \alpha_{D}\left(\gt_{\mba \mbb \mbe \mbf}\gt_{\mbe \mbf \mbc \mbd} + 2 \textrm{ terms} \right)  + \alpha_{S}\left(\gt_{\mba \mbb \mbe \mbf}\gt_{\mbe \mbg \mbh \mbc}\gt_{\mbf \mbg \mbh \mbd}+ 5 \textrm{ terms}\right) \crcr 
& \quad  \, + \, \alpha_{U} (\gt_{\mba \mbe \mbf \mbg} \gt_{\mbb \mbe \mbf \mbh} \gt_{\mbg \mbm \mbn \mbc} \gt_{\mbh \mbm \mbn \mbd} + 5 \textrm{ terms} )  + \, \alpha_{T} (\gt_{\mba \mbb \mbe \mbf} \gt_{\mbe \mbg \mbh \mbm} \gt_{\mbf \mbg \mbh \mbn} \gt_{\mbm \mbn \mbc \mbd} + 2 \textrm{ terms} )   \crcr
& \quad  \,  + \, \alpha_{I_1} (\gt_{\mba \mbb \mbe \mbf} \gt_{\mbe \mbg \mbh \mbm} \gt_{\mbf \mbg \mbn \mbc} \gt_{\mbh \mbm \mbn \mbd} + 11 \textrm{ terms} ) + \,\alpha_{I_2}(\gt_{\mba \mbb \mbe \mbf} \gt_{\mbe \mbg \mbh \mbc} \gt_{\mbf \mbm \mbn \mbd} \gt_{\mbg \mbh \mbm \mbn} + 5 \textrm{ terms} )  \crcr
& \quad  + \, \alpha_{I_3} (\gt_{\mba \mbb \mbe \mbf}\gt_{\mbh \mbm \mbn \mbf}\gt_{\mbh \mbm \mbn \mbg}\gt_{\mbg \mbe \mbc \mbd} +2 \textrm{ terms} ) + \, \alpha_{I_4} ( \gt_{\mba \mbe \mbm \mbh}\gt_{\mbb \mbe \mbf \mbn}\gt_{\mbc \mbf \mbm \mbg}\gt_{\mbd \mbg \mbn \mbh} ) 
\label{eq:beta_abcd_alpha}
\end{align}
\begin{align}
\beta^{(2)}_{\mbc\mbd}
&= - (d-2\Delta_{\phi} ) \rt_{\mbc \mbd} +\alpha_{D}  \big( \rt_{ \mbe \mbf}\gt_{\mbe \mbf \mbc \mbd} \big) + \alpha_{S} \big(\rt_{\mbe \mbf} \gt_{\mbe \mbg \mbh \mbc}\gt_{\mbf \mbg \mbh \mbd} \big) + \, \alpha_{T}  
\big(\rt_{\mbe \mbf} \gt_{\mbe \mbg \mbh \mbm} \gt_{\mbf \mbg \mbh \mbn} \gt_{\mbm \mbn \mbc \mbd} \big)  \crcr
& \quad 
 + \, \alpha_{I_1}
(\rt_{\mbe \mbf} \gt_{\mbe \mbg \mbh \mbm} \gt_{\mbf \mbg \mbn \mbc} \gt_{\mbh \mbm \mbn \mbd} + 1 \textrm{ term} )  + \, \alpha_{I_2}
\big(\rt_{ \mbe \mbf} \gt_{\mbe \mbg \mbh \mbc} \gt_{\mbf \mbm \mbn \mbd} \gt_{\mbg \mbh \mbm \mbn} \big)  \crcr 
& \quad + \, \alpha_{I_3}
\big( \rt_{\mbe \mbf} \gt_{\mbh \mbm \mbn \mbf} \gt_{\mbh \mbm \mbn \mbg} \gt_{\mbg \mbe \mbc \mbd} \big) \,. 
\label{eq:beta2_abcd_alpha}
\end{align}
The main result of this paper is the determination of the constants $\alpha$ in appendix \ref{app:integrals}:
\begin{align}
& \alpha_{D} \, = \, 1 +\frac{\epsilon}{2}\big[\psi(1)-\psi(\tfrac{d}{2}) \big]+\frac{\epsilon^2}{8}\left[\left(\psi(1)-\psi(\tfrac{d}{2})\right)^2+ \psi_1(1)-\psi_1(\tfrac{d}{2})\right] \,, \crcr
&\alpha_{S} \, = \,  2\psi( \tfrac{d}{4} ) - \psi( \tfrac{d}{2})-\psi(1)   +\frac{\epsilon}{4}\Big[\left[2\psi(\tfrac{d}{4})-\psi(\tfrac{d}{2})-\psi(1)\right]
\left[3\psi(1)-5\psi(\tfrac{d}{2})+2\psi(\tfrac{d}{4})\right]   \crcr
& \qquad \; + 3\psi_1(1) + 4\psi_1(\tfrac{d}{4})-7\psi_1(\tfrac{d}{2})  -4 J_0(\tfrac{d}{4}) \Big] \, , \crcr
& \alpha_{U} = \alpha_{I_2} =\,  - \psi_1(1)-\psi_1(\tfrac{d}{4})+2\psi_1( \tfrac{d}{2})
+ J_0(\tfrac{d}{4})\, , \crcr
& \alpha_{T} \, = \, \frac{1}{2}\Big[2\psi(\tfrac{d}{4}) - \psi(\tfrac{d}{2})-\psi(1) \Big]^2 + \frac{1}{2}  \psi_1(1)+ \psi_1(\tfrac{d}{4}) - \frac{3}{2} \psi_1(\tfrac{d}{2}) - \, J_0(\tfrac{d}{4}) \, , \crcr
& \alpha_{I_1} \, = \, \frac{3}{2}\left[2\psi( \tfrac{d}{4} ) - \psi( \tfrac{d}{2})-\psi(1)\right]^2
 + \frac{1}{2} \psi_1(1) 
-\frac{1}{2}\psi_1(\tfrac{d}{2})
\crcr
& \alpha_{I_3} \, = \,\frac{\Gamma(-\tfrac{d}{4})\Gamma(\tfrac{d}{2})^2}{3 \, \Gamma( \tfrac{3d}{4})}\, , 
\crcr
& \alpha_{I_4} \, = \,\frac{\, \Gamma(1 + \tfrac{d}{4})^3\Gamma(- \tfrac{d}{4})}{
 \, \Gamma(\tfrac{d}{2} )} \; 6 \, \Big[  \psi_1(1) -  \psi_1(\tfrac{d}{4})  \Big]  \,,
 \label{eq:alphas}
\end{align}
with $\psi_i$ the polygamma functions of order $i$ and $J_0$ the sum:
\begin{equation}
J_0(\tfrac{d}{4})=\frac{1}{\Gamma(\tfrac{d}{4})^2}\sum_{n \geq 1}\frac{\Gamma(n+\tfrac{d}{2})\Gamma(n+ \tfrac{d}{4})^2}{n(n!)\Gamma(\tfrac{d}{2}+2n)}\Big[2\psi(n+1)-\psi(n)-2\psi(n+\tfrac{d}{4})-\psi(n+\tfrac{d}{2})+2\psi(\tfrac{d}{2}+2n)\Big] \,.
\end{equation}
As we are interested in Wilson-Fisher-like fixed points, with $\gt$ of order $\epsilon$, we have expanded the constants $\alpha$ in $\epsilon$ up to a consistent order (such that the beta functions are written up to order $\epsilon^p \gt^q$ with $p+q= 4$).

Notice that $\alpha_{I_3} $ is the only constant blowing up for $d\to 4$. This is because it contains a (melonic/sunset) two-point subgraph that is finite for $d<4$, and thus not renormalized by a counterterm; however, in $d=4$ such subgraph is divergent, hence the singularity in $\alpha_{I_3} $.

\section{Applications}
\label{sec:appl}
The most general model \eqref{eq:action} has only $\mathbb{Z}_2$ symmetry, that is, invariance under simultaneous sign flip of all the fields.
However, we do not know how to solve the fixed point equations in full generality, and all the known interacting fixed points with $\cN\geq 2$ have a symmetry group strictly larger than $\mathbb{Z}_2$ \cite{Rychkov:2018vya}.
In this section, using the results obtained in the previous section, we will study the fixed-points of some specific models, characterized by their invariance under different symmetry groups. In particular we will only consider some of the models  which have been studied the most, at least in their short-range version, because of their physical interest.

\subsection{The long-range Ising model}
\label{sec:ising model}
The Ising model is the special case $N=1$ of the $O(N)$ vector model which we will discuss in the next subsection. We prefer to  discuss it separately
 in this subsection because of its physics and historical importance, and because it is the long-range model for which more results are available in the literature.

Setting $\cN=1$, $\gt_{\mba \mbb \mbc \mbd} =  \gt $ and $\rt_{\mba \mbb}=\rt$ in  \eqref{eq:beta_abcd_alpha} and \eqref{eq:beta2_abcd_alpha},  we find the beta functions:
	\begin{align}
		\beta^{(4)} \, &= \, - \epsilon \, \gt \, + \, 3 \, \alpha_D \, \gt^2 \, + \, 6 \, \alpha_S \, \gt^3
		\, + \, (3\alpha_T + 6 \alpha_U + 12\alpha_{I_1} + 6 \alpha_{I_2} + 3\alpha_{I_3} + \alpha_{I_4}) \gt^4 \, , \\
		\beta^{(2)} \, &= \, - (d-2\Delta_{\phi} ) \, \rt \, + \, \alpha_{D} \, \rt \, \gt \, + \, \alpha_{S} \, \rt \, \gt^2 \, + \, (\alpha_{T} + 2\alpha_{I_1} + \alpha_{I_2} + \alpha_{I_3}) \, \rt \, \gt^3  \, .
	\end{align}
Parametrizing the coefficients of the $\epsilon$ expansion  of the one- and two-loop constants $\alpha$ as: 
	\begin{align}
		\alpha_D \, &= \, 1\, + \, \alpha_{D,1} \, \epsilon \, + \, \alpha_{D,2} \, \epsilon^2 \, + \, \mathcal{O}(\epsilon^3) \, ,\crcr
		\alpha_S \, &= \, \alpha_{S,0} \, + \, \alpha_{S,1} \, \epsilon \,  + \, \mathcal{O}(\epsilon^2) \, ,
	\label{eq:alpha param}
	\end{align}
we can solve for the fixed point coupling perturbatively in $\epsilon$, obtaining $\rt_\star=0$ and:
	\begin{align}
		\gt_{\star} \, &= \, \frac{\epsilon}{3} \, - \, \left( \frac{3\alpha_{D,1} + 2\alpha_{S,0}}{9} \right) \, \epsilon^2
		\, - \, \frac{\epsilon^3}{81} \bigg[ 27 \alpha_{D,2} -24 \alpha_{S,0}^2 + 9 (2\alpha_{S,1} - 3\alpha_{D,1}^2) \nonumber\\
		&\hspace{120pt} - 54\alpha_{D,1} \alpha_{S,0} + 3\alpha_T + 6 \alpha_U + 12\alpha_{I_1} + 6 \alpha_{I_2} + 3\alpha_{I_3} + \alpha_{I_4} \bigg]
		\, + \, \mathcal{O}(\epsilon^4) \, .
	\end{align}
As the stability matrix is triangular, the stability exponents are simply given by:
\begin{align}
		\partial_{\gt}\beta^{(4)}(\gt_{\star}) \, &= \, \epsilon \, + \, \frac{2\alpha_{S,0}}{3} \, \epsilon^2
		\, + \, \frac{2\epsilon^3}{27} \bigg[ -12 \alpha_{S,0}^2 + 9 \alpha_{S,1} - 18\alpha_{D,1} \alpha_{S,0}\nonumber\\
		&\hspace{120pt} + 3\alpha_T + 6 \alpha_U + 12\alpha_{I_1} + 6 \alpha_{I_2} + 3\alpha_{I_3} + \alpha_{I_4} \bigg]
		\, + \, \mathcal{O}(\epsilon^4) \, , \\
		\partial_{\rt}\beta^{(2)}(\gt_{\star}) \, &= \, - (d-2\Delta_{\phi} ) \, + \, \frac{\epsilon}{3} \, - \, \frac{\alpha_{S,0}}{9} \, \epsilon^2 \nonumber\\
		&\quad \, - \, \frac{\epsilon^3}{81} \bigg[ -12 \alpha_{S,0}^2 + 9 \alpha_{S,1} 
		- 18 \alpha_{D,1} \alpha_{S,0} + 6 \alpha_U + 6\alpha_{I_1} + 3 \alpha_{I_2} + \alpha_{I_4} \bigg] \, + \, \mathcal{O}(\epsilon^4) \, .
\end{align}

As usual, the stability exponents are related to the critical exponents, which describe universal properties of  critical phenomena. The anomalous dimension $\eta$, the susceptibility exponent $\gamma$, and the correlation length exponent $\nu$ satisfy the scaling relation $\gamma= (2-\eta) \nu$, and  for the long-range models we have also $2-\eta = 2\zeta$.
Therefore, it suffices to consider $\nu$, which is given by:
	\begin{align}
		\nu^{-1} \, = \, - \, \partial_{\rt}\beta^{(2)}(\gt_{\star}) \, ,
	\label{eq:nu}
	\end{align}
and the correction-to-scaling exponent $\omega$, given by:
\be
\omega = \partial_{\gt}\beta^{(4)}(\gt_{\star}) \,.
\ee
Using \eqref{eq:alphas} and $d-2\Delta_{\phi}=(d+\epsilon)/2=2\zeta$, the exponents are obtained as:
\begin{align}
	\omega\, &= \, \epsilon \,+ \, \frac{2}{3} \epsilon ^2   \Big[ 2\psi(\tfrac{d}{4}) - \psi(\tfrac{d}{2}) -\psi(1) \Big] \crcr
	& \quad +\, \frac{1}{18} \epsilon ^3 \bigg[ 13 \left(
   2\psi(\tfrac{d}{4})-\psi(1)  -\psi(\tfrac{d}{2})\right)^2 +3\left(\psi_1(1)- \psi_1(\tfrac{d}{2})\right) \crcr
   & \qquad  + \frac{8 \Gamma \left(-\frac{d}{4}\right) \left(\Gamma \left(\frac{d}{2}\right)^3+\Gamma \left(\frac{3
   d}{4}\right) \Gamma \left(\frac{d}{4}+1\right)^3 \left(\psi_1(1)- \psi_1(\tfrac{d}{4})\right)\right)}{\Gamma
   \left(\frac{d}{2}\right) \Gamma \left(\frac{3 d}{4}\right)} \bigg] \, + \, \mathcal{O}(\epsilon^4) \,, 
    \label{eq:omega-Ising}\\
		\nu^{-1} \, &= \, \frac{d}{2} \, + \, \frac{\epsilon}{6} \, + \, \frac{\epsilon^2}{9} \Big[ 2\psi(\tfrac{d}{4}) - \psi(\tfrac{d}{2}) -\psi(1) \Big] \,  \nonumber\\
		&\quad + \, \frac{\epsilon^3}{108} \bigg[\psi_1(1)-\left(
   2\psi(\tfrac{d}{4})-\psi(1)  -\psi(\tfrac{d}{2})\right)^2-\psi_1(\tfrac{d}{2})\crcr 
		& \qquad +8\frac{\Gamma(1+\tfrac{d}{4})^3\Gamma(-\tfrac{d}{4})\left(\psi_1(1)-\psi_1(\tfrac{d}{4})\right)}{\Gamma(d/2)}\bigg] \, + \, \mathcal{O}(\epsilon^4) \, .
  \label{eq:nu-Ising}
\end{align}

Notice that $\alpha_{I_3}$ does not contribute to $\nu$, but it appears in $\omega$.
As mentioned earlier, $\alpha_{I_3}\to-\infty$ as we approach $d=4$, and therefore, while the coefficients of the $\epsilon$ expansion of $\nu$ remain small at this order, the three-loop coefficient for the dimension of the quartic operator becomes large in this limit. It has been argued in \cite{Honkonen:1990} that such growing coefficients provide an explanation of the transition from long-range to short-range behavior, which is supposed to happen when $2\zeta = 2-\eta_{\rm SR}$, where $\eta_{\rm SR}$ stands for the anomalous dimension of the short-range Ising model \cite{Sak:1973}.

In integer dimensions, $\nu^{-1}$ evaluates to:\footnote{We stress again that while short-range models cannot undergo a phase transition at $d=1$, the existence of a phase transition in the one-dimensional long-range Ising model with $0< 2\z < 1$ has been proved rigorously in \cite{Dyson:1968up}.}
\begin{align}
\nu^{-1} \, & = \, 1.5 + 0.1667\, \epsilon - 0.1812\, \epsilon^2 +  0.2633\, \epsilon^3 \,, & \text{at } \; d=3\,, \crcr
\nu^{-1} \, & = \,  1 + 0.1667\, \epsilon - 0.3081\, \epsilon^2 +  0.5301\, \epsilon^3 \,, & \text{at } \; d=2\,, \crcr
\nu^{-1} \, & = \,  0.5 + 0.1667\, \epsilon- 0.6571\, \epsilon^2 + 2.018\, \epsilon^3 \,, & \text{at } \; d=1\,.
\end{align}
In Table~\ref{tab:critexp} we compare the numerical values of $\nu$ in one dimension at different values of $\epsilon$ obtained from our loop expansion with those from numerical simulations \cite{Glumac:1989,Luijten:1997-thesis,uzelac2001critical,tomita2009monte}. Other numerical results can also be found in \cite{Luijten:1997} for the mean field regime, and in \cite{Loscar:2018,rodriguez2009study,Xu:1993} for other values of $\z$ in one and two dimensions.
As the perturbative series is only asymptotic, a summation method must be employed for its quantitative use (as for example in \cite{Guida:1998bx}). We have used the most basic method, the Pad\'e-Borel summation (e.g.\ \cite{Kleinert:2001ax} and references therein), with which we have found an improvement with respect to the naive series (which we report for comparison), especially at larger values of $\epsilon$.
In Table~\ref{tab:critexp-omega} we provide a similar comparison for $\omega$ in one dimension with available numerical simulations \cite{Luijten:1997-thesis}.
\begin{table}[htb]
\begin{center}
\begin{tabular}{|c|c||c|c|c|c|c|c|c|c|}\hline
$\epsilon$ &  $2\zeta$  &  mean-field      &     three-loop   &    PB $[2/1]$ &   Ref.~\cite{Glumac:1989}        &     Ref.~\cite{Luijten:1997-thesis}  & Ref.~\cite{uzelac2001critical} & Ref.~\cite{tomita2009monte}                  \\ \hline
0.2 & 0.6 &1.6667    &     1.9015    &  1.937(16)   &     2.16        &     1.98      & 2.00(16) & 1.80(21)    \\
0.4 & 0.7 & 1.4286   &      1.5559        &  1.94(6)  &    2.123       &      2.01 & 1.96(15) & 1.83(17)               \\
0.6 & 0.8 & 1.25   &    0.4881  &  1.98(12)  &     2.208        &       2.17 & 2.13(18) & 1.89(5)              \\
\hline
\end{tabular}\end{center}
\caption{The critical exponent $\nu$ for the long-range Ising model at $d=1$, as computed in mean-field ($\nu=1/(2\z)$), by the naive three-loop series, and by a Pad\'e-Borel (PB) summation with $[2/1]$ approximant. The error in the latter is estimated by the difference with the PB summation of the two-loop series with $[1/1]$ approximant. The last four columns report numerical results from the literature for comparison.
}
\label{tab:critexp}
\end{table}
\begin{table}[htb]
\begin{center}
\begin{tabular}{|c|c||c|c|c|c|}\hline
$\epsilon$ &  $2\zeta$ &   one-loop         &     three-loop    &    PB $[2/1]$       &     Ref.~\cite{Luijten:1997-thesis}                    \\ \hline
0.2 & 0.6 & 0.2          &     0.35    &  0.139(17)          &     0.15         \\
0.4 & 0.7 & 0.4      &     2.20      &  0.24(6)         &      0.23                 \\
0.6 & 0.8 & 0.6       &    7.39   &  0.33(11)       &       0.25              \\
\hline
\end{tabular}\end{center}
\caption{The critical exponent $\omega$ for the long-range Ising model at $d=1$, as computed by one- and three-loop truncations and by a Pad\'e-Borel summation of the three-loop series with $[2/1]$ approximant, with errors estimated as before. The last column reports numerical results from Ref.~\cite{Luijten:1997-thesis}  for comparison.
}
\label{tab:critexp-omega}
\end{table}

In Table \ref{tab:critexp2} we compare numerical values of $\nu$ in two dimensions at different values of $\epsilon$ obtained with our loop expansion and summation method with numerical simulations \cite{Angelini:2014}. The value $\epsilon=1.5$ is the maximum value of $\epsilon$ we consider because it corresponds to the transition between long-range and short-range behavior happening at $2\zeta=2-\eta_{SR}$ (and we indeed find a value of $\n$ consistent with the exact result in two dimensions, $\n_{SR}=1$). 

\begin{table}[htb]
\begin{center}
\begin{tabular}{|c|c||c|c|c|c|}\hline
$\epsilon$ &  $2\zeta$  & mean-field &     three-loop    &   PB $[2/1]$ &   Ref.~\cite{Angelini:2014}                         \\ \hline
0.4 & 1.2 &0.8333 &0.9463  & 0.966(6) &   0.977(34)     \\
0.61812 & 1.30906 & 0.7640& 0.8748 & 0.962(14) & 0.986(33)     \\
1.2 & 1.6 & 0.625& 0.1821 & 0.98(5) &  1.004(34)   \\
1.5 & 1.75 &0.5714 & -0.6457  & 0.99(7) &  1.02 (12) \\
\hline
\end{tabular}\end{center}
\caption{The critical exponent $\nu$ for the long-range Ising model at $d=2$, as computed in mean-field ($\nu=1/(2\z)$), by the naive three-loop series, and by a Pad\'e-Borel (PB) summation with $[2/1]$ approximant (with error estimated by the difference with the PB summation of the two-loop series with $[1/1]$ approximant). The last column reports numerical results from the literature for comparison.}
\label{tab:critexp2}
\end{table}

\paragraph{On the conjectured relation between short-range and long-range Ising models.}
The peculiar value $\epsilon=0.61812$ was considered in \cite{Angelini:2014} (and hence in our  Table \ref{tab:critexp2}) in order to test a conjecture relating the long-range Ising model at given dimension $d$, to the short-range Ising model at a different dimension $d_{SR}$ \cite{Banos:2012,Angelini:2014,Defenu:2014}. According to the conjecture, the effective dimension or effective $\z$ should be found from the relation
 \be \label{eq:LR-SR-1}
 \f{2\zeta}{d}= \f{2-\eta_{SR}(d_{SR})}{d_{SR}}\,,
 \ee
 and one should then also see other relations among the critical exponents, such as 
 \be \label{eq:LR-SR-2}
 d\,\nu(\z,d)=d_{SR}\, \nu_{SR}(d_{SR})\,, \quad \omega(\z,d)/d=\omega_{SR}(d_{SR})/d_{SR}\,.
 \ee
Therefore, the short-range model at $d_{SR}=3$ should be related to the long-range model at $d=2$ if one takes  $2\zeta=1.30906$, or equivalently $\epsilon=0.61812$.
It should be said that there is no compelling evidence in favor of this conjecture and actually we will now show that it is only an approximate relation valid close to the upper critical dimensions ($d_{SR}\simeq 4$ and $d\simeq 4\z$), as also observed numerically in \cite{Angelini:2014} and remarked in \cite{Paulos:2015jfa,Behan:2017emf}.
Close to the upper critical dimensions, equation \eqref{eq:LR-SR-1} can be solved perturbatively.
Consider indeed the long-range Ising model at arbitrary $d<4$ and with long-range exponent \eqref{eq:zeta-eps}, for small $\epsilon$; next, consider the short-range Ising model in $d_{SR}=4-\epsilon_{SR}$, for small $\epsilon_{SR}$. Using the known expansion of $\eta_{SR}(4-\epsilon_{SR})$ (e.g.\ \cite{Kleinert:2001ax}), from \eqref{eq:LR-SR-1} we obtain:
\be
\f{\epsilon}{d} = \f{\epsilon_{SR}}{4} + \f{23 \epsilon_{SR}^2}{432} + \f{ 185 \epsilon_{SR}^3}{46656} +\cO(\epsilon_{SR}^4) \,.
\ee
Plugging this expression into \eqref{eq:omega-Ising} and \eqref{eq:nu-Ising}, and using  \eqref{eq:LR-SR-2} in order to determine $\nu_{SR}(4-\epsilon_{SR})$ and $\omega_{SR}(4-\epsilon_{SR})$, we find that they agree with the known short-range exponents up to order $\epsilon_{SR}$, but they disagree already at order $\epsilon_{SR}^2$, with in particular some surviving polygamma functions that should not appear in the short-range case.
We conclude that the relations \eqref{eq:LR-SR-1} and \eqref{eq:LR-SR-2} can at best be approximate and qualitative, in particular they can be trusted as long as one can trust the one-loop approximation, for which they are exact.
In fact, from a numerical point of view, the results of the $d=2$ long-range Ising model with $\epsilon=0.61812$ and the $d_{SR}=3$ short-range one are quite close but not in perfect agreement.
Comparing the short-range results from the literature \cite{El-Showk:2014dwa}, $3\nu_{SR}(3)\simeq 1.88997(15)$, with our resummed result $2\nu_{LR}\simeq 1.924(28)$, we find that the values are approximately compatible with equation \eqref{eq:LR-SR-2}, within the given errors. 
On the other hand, from the Pad\'e-Borel summation of our three-loop result  at $d=2$ we find that $\omega/2\simeq 0.20(4)$, which is slightly off from  $\omega_{SR}(3)/3 \simeq 0.2767(6)$ obtained again from \cite{El-Showk:2014dwa}.

\subsection{The long-range $O(N)$ vector model}
\label{sec:vector model}

For general $\cN=N$, the largest possible symmetry group is $O(N)$, which reduces the number of couplings 
to just one, corresponding to the unique quartic invariant $(\phi_{\mba}\phi_{\mba})^2$.
The long-range quartic $O(N)$ model is defined by the choice of couplings:
	\begin{align} \label{eq:g-vector}
		\gt_{\mba \mbb \mbc \mbd} \, = \, \frac{\gt}{3} \, \big( \delta_{ab} \delta_{cd} + \delta_{ac} \delta_{bd} + \delta_{ad} \delta_{bc} \big) \, , \qquad
		\rt_{\mba \mbb} \, = \, \rt \, \delta_{ab}  \, ,
	\end{align}
where we have identified $\mba=a=1,\ldots ,N$, and so on.
By substituting \eqref{eq:g-vector} in \eqref{eq:beta_abcd_alpha} and \eqref{eq:beta2_abcd_alpha}, we find the beta functions:
	\begin{align}
		\beta^{(4)} \, &= \, - \epsilon \, \gt \, + \, \frac{\alpha_{D}}{3} (N+8) \gt^2 \, + \, \frac{2\alpha_{S}}{9} (5N + 22) \gt^3
		\,  \nonumber\\
		&\quad+ \, \Big[ (3N^2 +22N + 56)(2\alpha_{I_2}+ \alpha_{T}) + 2(N^2+20N+60)(2\alpha_{I_1}+ \alpha_{U})\nonumber\\
		&\qquad  + 3( N+8)(N+2) \alpha_{I_3} + (5N +22) \alpha_{I_4} \Big] \frac{\gt^4}{27} \, , \\
		\beta^{(2)} \, &= \, - (d-2\Delta_{\phi} ) \, \rt \, + \, \frac{\alpha_{D}}{3} \, (N+2) \, \rt \, \gt \, + \, \frac{\alpha_{S}}{3} \, (N+2) \, \rt \, \gt^2 \nonumber\\
		&\quad + \, \frac{(N+2)}{27} \, \Big( 3(N+2) (\alpha_{T} + \alpha_{I_3}) + (N+8) (2\alpha_{I_1} + \alpha_{I_2} ) \Big) \, \rt \, \gt^3  \, .
	\end{align}
Besides the trivial fixed point $\gt_{\star}=\rt_\star=0$, we also find a (perturbative in $\epsilon$) non-trivial fixed point with $\rt_\star=0$ and:
	\begin{align}
		\gt_{\star} \, &= \, \frac{3\epsilon}{N+8}
		\, - \, \frac{3\epsilon^2}{(N+8)^3} \, \Big[ (N+8)^2\alpha_{D,1} + 2(5N+22)\alpha_{S,0} \Big] \nonumber\\
		&\quad + \, \frac{3\epsilon^3}{(N+8)^5} \bigg[ (N+8)^2 \Big( (N+8)^2\alpha_{D,1}^2 - 2(5N +22)( \alpha_{S,1} -3 \alpha_{S,0} \alpha_{D,1})-3(N+2)\alpha_{I_3} \Big) \nonumber\\
		&\hspace{60pt} - (N+8) \Big( (3N^2 +22N + 56) (2\alpha_{I_2}+\alpha_{T}) + 2(N^2+20N+60)(2\alpha_{I_1}+ \alpha_{U})  \nonumber\\
		&\hspace{60pt}   + (5N +22) \alpha_{I_4} \Big)  -(N+8)^4 \alpha_{D,2} + 8(5N +22)^2 \alpha_{S,0}^2  \bigg]
		\, + \, \mathcal{O}(\epsilon^4) \, ,
	\end{align}
where we used the parametrization \eqref{eq:alpha param}.

This is the long-range version of the Wilson-Fisher fixed point, describing several important universality classes: self-avoiding walks ($N=0$), Ising  ($N=1$), XY  ($N=2$), Heisenberg  ($N=3$), and spherical  ($N=\infty$); see for example  \cite{Pelissetto:2000ek} for a review of their short-range version.

The long-range critical exponents are given by:
	\begin{align} \label{eq:omega-ON}
		\omega \equiv \partial_{\gt}\beta^{(4)}(\gt_{\star}) \, &= \, \epsilon \, + \, \frac{2(5N +22)\alpha_{S,0}}{(N+8)^2} \, \epsilon^2
		\, + \, \frac{2\epsilon^3}{(N+8)^4} \bigg[ -4 (5N +22)^2\alpha_{S,0}^2 \nonumber\\
	         & \qquad+ (N+8)^2 (5N +22) (\alpha_{S,1} -2 \alpha_{D,1}\alpha_{S,0}) \nonumber\\
		&\qquad + (N+8) \Big( (3N^2 +22N + 56)(2\alpha_{I_2}+ \alpha_{T}) + 2(N^2+20N+60)(2\alpha_{I_1}+ \alpha_{U} ) \nonumber\\
		&\hspace{70pt} + 3(N+8)(N+2) \alpha_{I_3} + (5N +22) \alpha_{I_4} \Big) \bigg]
		\, + \, \mathcal{O}(\epsilon^4)
		 \, , \\
		 \label{eq:nu-ON}
		\nu^{-1}\equiv - \partial_{\rt}\beta^{(2)}(\gt_{\star}) \, &= \, 
		 2\zeta \, - \, \frac{(N+2)}{N+8} \, \epsilon \, + \, \frac{(N+2)(7N+20)\alpha_{S,0}}{(N+8)^3} \, \epsilon^2 \nonumber\\
		&\quad + \, \frac{(N+2)\epsilon^3}{(N+8)^5} \bigg[ -4(5N+22)(7N+20) \alpha_{S,0}^2 + (N+8)^2(7N+20) (\alpha_{S,1} -2 \alpha_{D,1} \alpha_{S,0}) \nonumber\\
		&\qquad+ (N+8) \Big(-8(N-1) \alpha_{T} + 2(N^2+20N+60) \alpha_U + 2(N^2+24N+56) \alpha_{I_1} \nonumber\\
		&\hspace{70pt}+ (5N^2+28N+48) \alpha_{I_2} + (5N +22) \alpha_{I_4} \Big) \bigg] \, + \, \mathcal{O}(\epsilon^4)
		 \, . 
	\end{align}

We can compute the critical value of $N$, $N_c$, at which $\omega$ vanishes and the Wilson-Fisher fixed point becomes marginal. At first order in $\epsilon$, we find:
\begin{equation}
N_c=-8 \pm 6\sqrt{2|\alpha_{S,0}|}\epsilon^{1/2}+\mathcal{O}(\epsilon) \; .
\end{equation}
At small $\epsilon$ this corresponds to a negative value of $N$, hence for the bosonic model the quartic operator never becomes relevant. However,  it can cross marginality for symplectic fermions, whose perturbative series is related to the bosonic one by changing sign of $N$ (see for example the recent  \cite{Giuliani:2020aot}).

The two-loop results for $\nu$ and $\omega$ agree with those first reported in \cite{Fisher:1972zz} and \cite{Yamazaki:1977pt}, respectively, while the three-loop contributions are new. At low $N$ and integer $d$, $\nu^{-1}$ is:
\begin{align}
\nu^{-1}&=0.5+0.1\epsilon-0.8043\epsilon^2+2.002\epsilon^3 \,, \quad (d=1, N=2 ) \,, \crcr
\nu^{-1}&=1+0.1\epsilon-0.3771\epsilon^2+0.5306\epsilon^3 \,, \quad ( d=2, N=2 ) \,, \crcr
\nu^{-1}&=1.5+0.1\epsilon-0.2218\epsilon^2+0.2672\epsilon^3 \,, \quad ( d=3 , N=2 ) \,, \crcr
\nu^{-1}&=0.5+0.04545\epsilon-0.9109\epsilon^2+1.963\epsilon^3 \,, \quad ( d=1, N=3 ) \,, \crcr
\nu^{-1}&=1+0.04545\epsilon-0.4270\epsilon^2+0.5212\epsilon^3 \,, \quad ( d=2, N=3 ) \,, \crcr
\nu^{-1}&=1.5+0.04545\epsilon-0.2512\epsilon^2+0.2632\epsilon^3 \,, \quad ( d=3, N=3 ) \,.
\end{align}

In Table \ref{tab:critvect22} we report the numerical values of $\nu$ and $\omega$ in two dimensions at $N=2$ and in three dimensions at $N=2,3$ for different values of $\epsilon$. 

\begin{table}[htb]
\begin{center}
\begin{tabular}{|c|c|c||c|c|c| |c|c|c| }\hline
\multirow{2}{*}{$(d,N)$ } & \multirow{2}{*} {$\epsilon$} &  \multirow{2}{*}{$2\zeta$} &  \multicolumn{3}{c||}{$\nu$}  
& 
 \multicolumn{2}{c|}{ $\omega$ }\\
& & & mean-field  &     three-loop    &    PB $[2/1]$  & three-loop &  PB $[2/1]$\\ \hline
\multirow{3}{*}{(2,2)}
& 0.2 & 1.1 &0.9091     &     0.9906   &  0.992(4)  & 0.1945 & 0.160(8)\\
& 0.4 & 1.2 & 0.8333    &   0.9831      &  1.000(18) & 0.6399 & 0.287(35)\\
& 0.6 & 1.3 & 0.7692  &    0.9482   &  1.02(4)  & 1.729 & 0.40(7)\\
\hline
\multirow{3}{*}{(3,2)} 
& 0.2 & 1.6 &0.625  &     0.6608   &  0.6610(7)  & 0.1835 & 0.173(5) \\
& 0.4 & 1.7 & 0.5882 &      0.6567      &  0.6600(35)  & 0.4350 & 0.317(25)\\
& 0.6 & 1.8 & 0.5556  &    0.6480   &  0.662(8)   &  0.9061 & 0.45(6)\\
\hline
\multirow{3}{*}{(3,3)} 
& 0.2 & 1.6 &0.625    &     0.6661   &  0.6663(18)   & 0.1832 & 0.174(5)\\
& 0.4 & 1.7 & 0.5882 &      0.6686      &  0.671(8)  & 0.4251 & 0.319(24)\\
& 0.6 & 1.8 & 0.5556 &    0.6682   &  0.680(17)   & 0.8642  &  0.45(5) \\
\hline
\end{tabular}\end{center}
\caption{The critical exponents $\nu$ and $\omega$ for the long-range vector model at various $d$ and $N$ computed in mean-field ($\nu=1/(2\z)$), by the naive three-loop series, and by a Pad\'e-Borel (PB) summation with $[2/1]$ approximant (with error estimated by the difference with the PB summation of the two-loop series with $[1/1]$ approximant).}
\label{tab:critvect22}
\end{table}

\bigbreak

\paragraph{Large-$N$.}
Next, we consider the $1/N$ expansion, which can be obtained either by rescaling the coupling by $\gt \to \bar{g}/N$ and following the usual large-$N$ analysis from the beginning, or by expanding the finite-$N$ result \eqref{eq:omega-ON} and \eqref{eq:nu-ON}.
Up to order $\mathcal{O}(N^{-2})$, the critical exponents are given by:
	\begin{align}
		\omega \, &= \, \epsilon   \, + \, \frac{2\epsilon^2}{ N} \Big[ 5 \alpha_{S,0}
		+ (3\alpha_{T} + 2\alpha_{U} + 4\alpha_{I_1} + 6\alpha_{I_2} + 3 \alpha_{I_3}+5(\alpha_{S,1}-2\alpha_{D,1}\alpha_{S,0}) ) \epsilon \Big] \crcr
		&\quad - \frac{2\epsilon^2}{N^2}\Big[ 58\alpha_{S,0}+\big(16\alpha_{I_1}+100\alpha_{I_2}+42\alpha_{I_3}-5\alpha_{I_4}+58(\alpha_{S,1}-2\alpha_{D,1}\alpha_{S,0})\crcr
& \qquad \quad +100\alpha_{S,0}^2+50\alpha_T+8\alpha_U\big)\epsilon\Big] \, + \, \mathcal{O}(N^{-3},\epsilon^4)  \, , \\
		\nu^{-1}\, & = \, 2\Delta_{\phi}
		\, + \, \frac{\epsilon}{ N} \, \Big[ 6 + 7 \alpha_{S,0} \epsilon
		+ \big(2\alpha_{U} + 2\alpha_{I_1} + 5\alpha_{I_2}+7(\alpha_{S,1}-2\alpha_{D,1}\alpha_{S,0})\big) \epsilon^2\Big]\crcr
		& \quad - \frac{\epsilon}{N^2}\,\Big[48+134\alpha_{S,0}\epsilon +\big(12\alpha_{I_1}+122\alpha_{I_2}-5\alpha_{I_4}+134(\alpha_{S,1}-2\alpha_{D,1}\alpha_{S,0})\crcr
&\qquad \quad +140\alpha_{S,0}^2+8\alpha_{T}+20\alpha_U\big)\epsilon^2 \Big] \, + \, \mathcal{O}(N^{-3},\epsilon^4) \, .
	\end{align}
The leading-order term of $\nu^{-1}$ (where we used $2\z-\epsilon =d-2\Delta_{\phi} -\epsilon =  2\Delta_{\phi}$) gives the spherical model result $\g= 2\z \nu = 2\zeta/(d-2\z)$, obtained first in \cite{Joyce:1966}.
The order $1/N$ contribution to $\nu$ was computed to all orders in $\epsilon$ in  \cite{Fisher:1972zz} and we reproduce it up to three loops.
The order $1/N^2$ contribution is new.

\subsection{The long-range cubic model}
\label{sec:cubic}

The next model we consider is obtained by breaking explicitly the $O(N)$ symmetry with an interaction of the form $\sum_{\mba} \phi_{\mba}^4$.
As the four fields share the same index, the continuous symmetry is completely broken; however, we can still independently flip the sign of any component, or permute them. In other words, we are left with the (hyper-)cubic symmetry group $(\mathbb{Z}_2)^N \rtimes S_N$.
For reviews of the short-range model with the same symmetry group, see \cite{Aharony:1976br,Pelissetto:2000ek,Kleinert:2001ax}; for some of the most recent results, using renormalization or bootstrap methods, see \cite{Stergiou:2018gjj,Kousvos:2018rhl,Kousvos:2019hgc,Adzhemyan:2019gvv,Antipin:2019vdg} and references therein. 
The long-range version of the model has been mostly unexplored, and we are only aware of the two-loop calculations of \cite{Yamazaki:1978-cubic,Yamazaki:1981-cubic,Chen:2001}.

The model is defined by the following choice of coupling $g_{\mba\mbb\mbc\mbd}$:
\begin{equation}
g_{\mba\mbb\mbc\mbd}=\frac{g_d}{3}\left(\delta_{ab}\delta_{cd}+\delta_{ac}\delta_{bd}+\delta_{ad}\delta_{bc}\right)
+g_c \delta_{ab}\delta_{ac}\delta_{ad} \,.
\label{eq:cubic_coupling}
\end{equation}
Substituting \eqref{eq:cubic_coupling} into \eqref{eq:beta_abcd_alpha} and \eqref{eq:beta2_abcd_alpha}, we find the beta functions up to three loops:
\begin{align}
\beta^{(4)}_d= & -\epsilon \tilde{g}_d+\frac{\alpha_{D}}{3}\left[6\tilde{g}_c+(N+8)\tilde{g}_d\right]\tilde{g}_d+\frac{2\alpha_{S}}{9}\left[9\tilde{g}_c^2+36\tilde{g}_c\tilde{g}_d+(5N+22)\tilde{g}_d^2\right]\tilde{g}_d\crcr
&  +2\Big[2\alpha_{I_1}+\alpha_{I_2}+\alpha_{I_3}+\alpha_{T}\Big]\tilde{g}_d\tilde{g}_c^3 +\frac{1}{3}\Big[52\alpha_{I_1}+30\alpha_{I_2}+(N+20)\alpha_{I_3}\crcr
&  +(N+18)\alpha_{T}+2\alpha_{I_4}+16\alpha_{U}\Big]\tilde{g}_d^2\tilde{g_c}^2  + \frac{2}{9}\Big[4(2N+27)\alpha_{I_1}+(9N+52)\alpha_{I_2}\crcr
&  +6(N+5)\alpha_{I_3}+8\alpha_{I_4}+2(3N+13)\alpha_{T}+4(N+12)\alpha_{U}\Big]\tilde{g}_d^3\tilde{g}_c \crcr
&+\frac{1}{27}\Big[2(N^2+20N+60)(2\alpha_{I_1}+\alpha_{U})+(3N^2+22N+56)(2\alpha_{I_2}+\alpha_{T})\crcr
& \qquad +3(N+2)(N+8)\alpha_{I_3}+(5N+22)\alpha_{I_4}  \Big]\tilde{g}_d^4  \, ,\crcr
\beta^{(4)}_c=& -\epsilon \tilde{g}_c+\alpha_{D}\left[3\tilde{g}_c+4\tilde{g}_d\right]\tilde{g}_c+\frac{2\alpha_{S}}{3}\left[9\tilde{g}_c^2+24\tilde{g}_c\tilde{g}_d+(N+14)\tilde{g}_d^2\right]\tilde{g}_c \crcr
& +\Big[12\alpha_{I_1}+6\alpha_{I_2}+3\alpha_{I_3}+\alpha_{I_4}+3\alpha_{T}+6\alpha_{U}\Big]\tilde{g}_c^4\crcr
& +2\Big[22\alpha_{I_1}+11\alpha_{I_2}+5\alpha_{I_3}+2\alpha_{I_4}+5\alpha_{T}+12\alpha_{U}\Big]\tilde{g}_c^3\tilde{g}_d \crcr
& +\frac{1}{3}\Big[4(N+40)\alpha_{I_1}+6(N+12)\alpha_{I_2}+3(N+10)\alpha_{I_3}\crcr
& \qquad +16\alpha_{I_4}+(N+34)\alpha_{T}+4(N+22)\alpha_{U}\Big]\tilde{g}_c^2\tilde{g}_d^2 \crcr
&+\frac{2}{27}\Big[12(3N+22)\alpha_{I_1}+3(N^2+6N+40)\alpha_{I_2}+18(N+2)\alpha_{I_3}\crcr
& \qquad +2(N+14)\alpha_{I_4}+6(N+10)\alpha_{T}+24(N+6)\alpha_{U}\Big]\tilde{g}_c\tilde{g}_d^3 \, , \crcr
\beta^{(2)}=& -(d-2\Delta_{\phi})\tilde{r}+\alpha_{D}\Big[\tilde{g}_c+\frac{\tilde{g}_d}{3}(N+2)\Big]\tilde{r}+\frac{\alpha_{S}}{3}\Big[3\tilde{g}_c^2+6\tilde{g}_c\tilde{g}_d+(N+2)\tilde{g}_d^2\Big]\tilde{r}\crcr
&+\Big[2\alpha_{I_1}+\alpha_{I_2}+\alpha_{I_3}+\alpha_{T}\Big]\tilde{g}_c^3\tilde{r}+\frac{1}{3}\Big[18\alpha_{I_1}+9\alpha_{I_2}+(N+8)(\alpha_{I_3}+\alpha_{T})\Big]\tilde{g_c^2}\tilde{g}_d\tilde{r} \crcr
& +\Big[(N+8)(2\alpha_{I_1}+\alpha_{I_2})+3(N+2)(\alpha_{I_3}+\alpha_{T})\Big]\Big[\frac{N+2}{27}\tilde{g}_d+\frac{1}{3}\tilde{g}_c\Big]\tilde{g}_d^2\tilde{r} \, .
\end{align}

Using \eqref{eq:alpha param} and expanding in $\epsilon$, we find three non-trivial fixed points.

\begin{description}
 \item [Heisenberg fixed point.]
The first fixed point, with $g_c=0$, is the $O(N)$ symmetric Heisenberg fixed point:
\begin{align}
\tilde{g}_d^{\star,H}&=\frac{3\epsilon}{N+8}-\frac{3\epsilon^2}{(N+8)^3}\left[(N+8)^2\alpha_{D,1}+2(5N+22)\alpha_{S,0}\right] \crcr
&+\frac{3\epsilon^3}{(N+8)^4}\Big[-2(N^2+20N+60)(2\alpha_{I_1}+\alpha_{U})-(3N^2+22N+56)(2\alpha_{I_2}+\alpha_{T}) \crcr
&-3(N+8)(N+2)\alpha_{I_3}-(5N+22)\alpha_{I_4} +(N+8)^3(\alpha_{D,1}^2-\alpha_{D,2})\crcr
&-2(N+8)(5N+22)(\alpha_{S,1}-3\alpha_{D,1}\alpha_{S,0})+8\frac{(5N+22)^2}{N+8}\alpha_{S,0}^2\Big]+\mathcal{O}(\epsilon^4)\,,\crcr
\tilde{g}_c^{\star,H}&=0 \,. 
\end{align}
\item[Ising fixed point.] The second one, with $g_d=0$, is the Ising fixed point:
\begin{align}
\tilde{g}_d^{\star,I}&=0 \,, \crcr
 \tilde{g}_c^{\star,I}&=\frac{\epsilon}{3}-\frac{\epsilon^2}{9}\Big[3\alpha_{D,1}+2\alpha_{S,0}\Big] +\frac{\epsilon^3}{81}\Big[-3(4\alpha_{I_1}+2\alpha_{I_2}+\alpha_{I_3})-\alpha_{I_4}+27(\alpha_{D,1}^2-\alpha_{D,2}+2\alpha_{D,1}\alpha_{S,0})\crcr
 &+6(4\alpha_{S,0}^2-3\alpha_{S,1})-3(\alpha_{T}+2\alpha_{U})\Big]+ \mathcal{O}(\epsilon^4)  \,.
\end{align}
\item[Cubic fixed point.] The last fixed point, with both couplings non-zero, is the cubic fixed point:
\begin{align}
\tilde{g}_d^{\star,C}&=\frac{\epsilon}{N}+\frac{\epsilon^2}{3N^3}\Big[2(N-1)(N-6)\alpha_{S,0}-3N^2\alpha_{D,1}\Big]  \crcr
& -\frac{\epsilon^3}{27N^5}\Big[-12N(N^3-6N^2+2N+4)\alpha_{I_1}-6N(N^3-9N^2+14N-8)\alpha_{I_2}\crcr
& +3N^2(N-1)(N+2)\alpha_{I_3}-N(2N^3-6N^2-7N+14)\alpha_{I_4}-27N^4(\alpha_{D,1}^2-\alpha_{D,2})\crcr
&+18N^2(N-1)(N-6)(3\alpha_{D,1}\alpha_{S,0}-\alpha_{S,1})+24(N-1)(N^3+5N^2-40N+36)\alpha_{S,0}^2\crcr
& +3N(N^3+3N^2-10N+8)\alpha_{T}-6N(2N^3-9N^2+4N+4)\alpha_{U}\Big] + \mathcal{O}(\epsilon^4) \,, \crcr
 \tilde{g}_c^{\star,C}&=\frac{\epsilon(N-4)}{3N}-\frac{\epsilon^2}{9N^3}\left(3N^2(N-4)\alpha_{D,1}+2(N-1)(N^2+6N-24)\alpha_{S,0}\right) \crcr
 &-\frac{\epsilon^3}{81N^5}\Big[12N(N^2-2N-2)(N^2+6N-8)\alpha_{I_1}+6N(N^4+7N^3-46N^2+64N-32)\alpha_{I_2}\crcr
 & +3N^2(N-1)(N+2)(N-4)\alpha_{I_3}+N(N+2)(N^3+6N^2-32N+28)\alpha_{I_4}\crcr
 & -27N^4(N-4)(\alpha_{D,1}^2-\alpha_{D,2})-18N^2(N-1)(N^2+6N-24)(3\alpha_{D,1}\alpha_{S,0}-\alpha_{S,1})\crcr
 & -24(N-1)(N^4+5N^3+28N^2-172N+144)\alpha_{S,0}^2+3N(N^4-5N^3-10N^2+40N-32)\alpha_{T}\crcr
 & +6N(N^4+10N^3-40N^2+16N+16)\alpha_{U}\Big] + \mathcal{O}(\epsilon^4) \,. 
\end{align}
\end{description}

In the limit $N \rightarrow \infty$, the cubic and Ising fixed points are equal. For $N=2$, the cubic and Ising fixed points verify:
\begin{equation}
\tilde{g}_d^{\star,C}=\tilde{g}_d^{\star,I}+\frac{3}{2}\tilde{g}_c^{\star,I} \,, \; ~~ \tilde{g}_c^{\star,C}=-\tilde{g}_c^{\star,I} \, ,
\end{equation}
which was first noticed  in the short-range case in \cite{Kleinert:2001ax}.

The critical exponents are the eigenvalues of the stability matrix. As the latter has a block triangular structure, the correction-to-scaling exponents $\omega_d$ and $\omega_c$ are the eigenvalues of the reduced stability matrix $\partial (\beta_s, \beta_d)/\partial (\gt_s, \gt_d)|_{\gt=\gt_\star}$, while the correlation length exponent is simply obtained from $\partial_{\rt} \beta^{(2)}|_{\gt=\gt_\star,\rt=0}$. 
They are given by the following expressions for the three non-trivial fixed points.

\begin{description}
 \item [Heisenberg fixed point.] For the Heisenberg fixed point we have:
\begin{align}
\omega_d^H=\, &\epsilon+\frac{2\epsilon^2(5N+22)\alpha_{S,0}}{(N+8)^2}\crcr
&+\frac{2\epsilon^3}{(N+8)^3}\Big[2(N^2+20N+60)(2\alpha_{I_1}+\alpha_{U})+(3N^2+22N+56)(2\alpha_{I_2}+\alpha_{T})\crcr
&\quad+3(N+8)(N+2)\alpha_{I_3}+(5N+22)\alpha_{I_4}-(N+8)(5N+22)(2\alpha_{D,1}\alpha_{S,0}-\alpha_{S,1})\crcr
&\quad-4\frac{(5N+22)^2}{N+8}\alpha_{S,0}^2\Big]+\mathcal{O}(\epsilon^4)\,, \crcr
\omega_c^H=\, &-\epsilon\frac{N-4}{N+8}+\frac{6\epsilon^2(N^2+2N+24)\alpha_{S,0}}{(N+8)^3} \crcr
& +\frac{2\epsilon^3}{(N+8)^4}\Big[12(N^2+6N+56)\alpha_{I_1}+3(N^3+2N^2+96)\alpha_{I_2}+2(N^2+7N+46)\alpha_{I_4}\crcr 
&\quad-12\frac{(5N+22)(N^2+2N+24)}{N+8}\alpha_{S,0}^2+3(N+8)(N^2+2N+24)(\alpha_{S,1}-2\alpha_{D,1}\alpha_{S,0})\crcr
&\quad-12(N^2+2N-12)\alpha_{T}+12(N^2+8N+36)\alpha_{U}\Big] +\mathcal{O}(\epsilon^4) \,, \crcr
\nu_H^{-1}=\, & 2\zeta 
-\epsilon\frac{N+2}{N+8}+\epsilon^2\frac{(N+2)(7N+20)\alpha_{S,0}}{(N+8)^3}\crcr
&-\frac{\epsilon^3(N+2)}{(N+8)^4}\Big[2(N^2+24N+56)\alpha_{I_1}+(5N^2+28N+48)\alpha_{I_2}+(5N+22)\alpha_{I_4}\crcr
&\quad -\frac{4(5N+22)(7N+20)}{N+8}\alpha_{S,0}^2+(N+8)(7N+20)(\alpha_{S,1}-2\alpha_{D,1}\alpha_{S,0})-8(N-1)\alpha_{T}\crcr
&\quad +2(N^2+20N+60)\alpha_{U} \Big] +\mathcal{O}(\epsilon^4)\,.  
\end{align}
\item[Ising fixed point.] For the Ising fixed point we have:
\begin{align}
\omega_d^I=\, &-\frac{\epsilon}{3}-\frac{2\epsilon^2\alpha_{S,0}}{9}-\frac{2\epsilon^3}{81}\Big[6\alpha_{I_1}+3\alpha_{I_2}+\alpha_{I_4}-6\alpha_{S,0}(3\alpha_{D,1}+2\alpha_{S,0})\crcr
& \qquad +9\alpha_{S,1}+6\alpha_{U}\Big]+\mathcal{O}(\epsilon^4) \,, \crcr 
\omega_c^I=\, &\epsilon+\frac{2\epsilon^2\alpha_{S,0}}{3} +\frac{2\epsilon^3}{27}\Big[12\alpha_{I_1}+6\alpha_{I_2}+3\alpha_{I_3}+\alpha_{I_4}-6\alpha_{S,0}(3\alpha_{D,1}+2\alpha_{S,0})\crcr
& \qquad +9\alpha_{S,1}+3\alpha_{T}+6\alpha_{U}\Big] +\mathcal{O}(\epsilon^4) \,,\crcr 
\nu_I^{-1}=\, & 2\zeta 
-\frac{\epsilon}{3}+\epsilon^2\frac{\alpha_{S,0}}{9}-\frac{\epsilon^3}{81}\Big[-6\alpha_{I_1}-3\alpha_{I_2}-\alpha_{I_4} +6\alpha_{S,0}(3\alpha_{D,1}+2\alpha_{S,0})\crcr
& \qquad -9\alpha_{S,1}-6\alpha_{U}\Big] +\mathcal{O}(\epsilon^4) \,. 
\end{align}
\item[Cubic fixed point.] For the cubic fixed point we have:
\begin{align}
\omega_d^C=\, &\epsilon+\frac{2\epsilon^2(N-1)(N^2+12)\alpha_{S,0}}{3N^2(N+2)}\crcr
& +\frac{2\epsilon^3}{27N^2(N+2)}\Big[12(N^4-2N^3+14N^2-8N-8)\alpha_{I_1}+6(N^4+N^3+8N^2-20N+16)\alpha_{I_2}\crcr
&\quad +3N(N+2)^2(N-1)\alpha_{I_3}+(N^4-4N^3+16N^2+6N-28)\alpha_{I_4}\crcr
&\quad+9N(N-1)(N^2+12)(\alpha_{S,1}-2\alpha_{D,1}\alpha_{S,0})\crcr
&\quad -12\frac{(N-1)(N^6+8N^5-14N^4+4N^3+384N^2-176N-288)}{N(N+2)^2}\alpha_{S,0}^2\crcr
&\quad +3(N^4+N^3+8N^2-20N+16)\alpha_{T} +6(N^4-2N^3+14N^2-8N-8)\alpha_{U}\Big]+\mathcal{O}(\epsilon^4) \,,\crcr
\omega_c^C=\, &\epsilon\frac{N-4}{3N}+\frac{2\epsilon^2(N-1)(N^3-4N^2-36N+48)\alpha_{S,0}}{9N^3(N+2)} \crcr
& +\frac{2(N-1)\epsilon^3}{81N^4(N+2)}\Big[6(N^4+2N^3-68N^2+72N+32)\alpha_{I_1}\crcr
&\quad +3(N^4-4N^3-44N^2+96N-64)\alpha_{I_2} +(N^4-N^3-44N^2+24N+56)\alpha_{I_4}\crcr
&\quad +9N(N^3-4N^2-36N+48)(\alpha_{S,1}-2\alpha_{D,1}\alpha_{S,0})\crcr
&\quad -12\frac{N^7+22N^5-180N^4-864N^3+1616N^2+800N-1152}{N(N+2)^2}\alpha_{S,0}^2\crcr
&\quad +12(N-2)(N^2-2N+4)\alpha_{T} +6(N^4-2N^3-40N^2+40N+16)\alpha_{U}\Big]+\mathcal{O}(\epsilon^4) \,,\crcr
\nu_C^{-1}=\, & 2\zeta 
-2\epsilon\frac{N-1}{3N}-\epsilon^2\frac{(N-1)(N^2-18N+24)\alpha_{S,0}}{9N^3}\crcr
& -\frac{\epsilon^3(N-1)}{81N^5}\Big[6N(N^3-14N^2+8N+16)\alpha_{I_1}+3N(N^3-26N^2+56N-32)\alpha_{I_2}\crcr
&\quad +N(N+2)(N^2-11N+14)\alpha_{I_4}+9N^2(N^2-18N+24)(\alpha_{S,1}-2\alpha_{S,0}\alpha_{D,1})\crcr
&\quad -12(N-1)(N^3+4N^2-100N+144)\alpha_{S,0}^2+24N(N-2)\alpha_{T}\crcr
&\quad +6N(N^3-14N^2+12N+8)\alpha_{U}\Big]+\mathcal{O}(\epsilon^4) \,,
\end{align}
with which we recover at two-loop the results of \cite{Yamazaki:1978-cubic,Yamazaki:1981-cubic,Chen:2001}.

\end{description}

The Gaussian fixed point is doubly unstable, while at the Ising fixed-point one eigenvalue is negative ($\omega_d^I$) and the other is positive. 

The stability of the Heisenberg and cubic fixed points are related. Indeed, there exists a critical value $N=N_c$ for which the cubic and Heisenberg fixed points collapse and exchange stability, as in the short-range model \cite{Kleinert:2001ax}. The Heisenberg fixed point is stable for $N<N_c$ and the cubic fixed point is stable for  $N>N_c$. Using the condition $\tilde{g}_c^{\star,C}=0$ (or equivalently $\omega_c^C=0$), we find the following $\epsilon$ expansion for $N_c$:
\begin{align}
N_c&= 4+2\epsilon\alpha_{S,0}+\frac{\epsilon^2}{6}\left(8\alpha_{I_1}+4\alpha_{I_2}+\frac{5}{4}\alpha_{I_4}+12(\alpha_{S,1}-2\alpha_{D,1}\alpha_{S,0})-13\alpha_{S,0}^2-\alpha_{T}+7\alpha_{U}\right) \,. 
\end{align}
We notice that the two-loop result (order $\epsilon$) coincides with the similar short-range result \cite{Kleinert:2001ax} upon taking $\alpha_{S,0}\to -1$, which corresponds to the value at $d=4$. The three-loop results are instead not related in a similar way.

Physically, the value $N=3$ is very interesting because the $O(3)$-symmetric fixed point characterizes the critical behavior of the Heisenberg model of magnetism. It is thus important to know whether $N_c$ is below or above $3$. Indeed, if $N_c$ is greater than $3$ all magnetic systems with cubic symmetry will have a $O(3)$-symmetric critical behavior as the Heisenberg fixed point will be the relevant one. However, if $N_c$ is smaller than $3$ the cubic fixed point will be the relevant one and will govern the critical behavior of magnetic systems with cubic symmetry. Table \ref{tab:nc} gives values of $N_c$ at $d=3$ for different values of $\epsilon$ obtained at different loop-orders and with the Pad\'e-Borel summation method. 
For $\epsilon=0.2,0.4$ we find $N_c$ greater than $3$ whereas it is smaller than $3$ for $\epsilon=0.6$. However, for short-range cubic models, results at three loops (in $d=4-\epsilon_{SR}$, extrapolated to $\epsilon_{SR}=1$) gave $N_c$ above $3$ while higher loop computations \cite{Adzhemyan:2019gvv,Kleinert:2001ax} gave values of $N_c$ below three. In order to accurately conclude on the value of $N_c$ in the long-range model for $\epsilon$ close to 1 
we need higher-loop results.

\begin{table}[htb]
\begin{center}
\begin{tabular}{|c||c|c|c|}\hline
$\epsilon$ &   one-loop          &     three-loop    &    PB $[1/1]$  \\ \hline
0.2 & 4 &        3.5712   &  3.500(5)  \\
 0.4 & 4 &      3.5897      &  3.171(13)                 \\
 0.6 & 4 &      4.0553  &  2.926(21)   \\
\hline
\end{tabular}\end{center}
\caption{The critical value $N_c$ for the long-range cubic model at $d=3$, as computed by a one-loop and three-loop truncation and by a Pad\'e-Borel summation of the three-loop series with $[1/1]$ approximant (the error is estimated by the difference with the PB summation of the two-loop series with $[0/1]$ approximant).}
\label{tab:nc}
\end{table}

We can nonetheless study the numerics of the critical exponents $\nu_C$ and  $\nu_H$ of the cubic and Heisenberg fixed points at $N=3$ for different values of $\epsilon$. The exponent $\nu_H$ at $d=3,N=3$ is identical with the exponent $\nu$ reported in table \ref{tab:critvect22}. For comparison, the corresponding $\nu_C$ is displayed in table \ref{tab:nucubic}. 

\begin{table}[h]
\begin{center}
\begin{tabular}{|c|c||c|c|c|}\hline
$\epsilon$& $2\zeta$ &  mean-field &   three-loop     &   PB $[2/1]$      \\ \hline
0.2 & 1.6 &   0.625 &     0.6657   &  0.6659(20)   \\
0.4 & 1.7 & 0.5882  &      0.6681      &  0.671(8)                 \\
0.6 & 1.8 &  0.5556 &    0.6674   &  0.681(19)   \\
\hline
\end{tabular}\end{center}
\caption{The critical exponents $\nu_C$  for the long-range cubic model at $d=3$ and $N=3$, as computed by a one-loop and three-loop truncation and by a Pad\'e-Borel summation of the three-loop series with $[2/1]$ approximant (with error estimated by the difference with the PB summation of the two-loop series with $[1/1]$ approximant).}
\label{tab:nucubic}
\end{table}

The results of the Pad\'e-Borel summation show that $\nu_H$ and $\nu_C$ lie very close to each other. This is due to the fact that $N_c\simeq 3$, so that for $N=3$ the two fixed points are very close to each other. As a consequence, for the physically interesting case $N=3$, the Heisenberg and the cubic critical behavior are practically indistinguishable, as noticed in \cite{Kleinert:1996hy}.

\subsection{The long-range $O(M) \times O(N)$ bifundamental model}
\label{sec:bifundamental}

The last special case of the general multi-scalar model we discuss is the $O(M)\times O(N)$ model. We consider two integers $M$ and $N$, such that $\cN=MN$, and we impose $O(M)\times O(N)$ symmetry.\footnote{In order to have a faithful action of the symmetry group, we should take the latter to be $O(M)\times O(N)/\mathbb{Z}_2$. In the rest of the paper, for conciseness we will simply refer to $O(M)\times O(N)$ symmetry, as very common in the literature.}
The resulting model, called \emph{bifundamental model} in \cite{Rychkov:2018vya}, has two quartic couplings, and it has been extensively studied in its short-range version (e.g.\ \cite{Kawamura:1988,Kawamura:1990,Pelissetto:2001fi,Gracey:2002pm,Delamotte:2003dw,Kompaniets:2020,Henriksson:2020fqi}).
However, we are not aware of any work on its long-range version.

The model is obtained by the substitution:
	\begin{align} \label{eq:g-bifund}
		g_{\mba \mbb \mbc \mbd} \, &= \, 
		\frac{g_s}{6} \big( \delta_{a_1 b_1} \delta_{c_1 d_1} ( \delta_{a_2 c_2}\delta_{b_2 d_2}+ \delta_{a_2 d_2}\delta_{b_2 c_2})\, + \, 
		 \delta_{a_1 c_1} \delta_{b_1 d_1} (\delta_{a_2 b_2} \delta_{c_2 d_2} +\delta_{a_2 d_2} \delta_{c_2 b_2})\crcr
		&\qquad\quad + \,
		 \delta_{a_1 d_1} \delta_{c_1 b_1} ( \delta_{a_2 c_2}\delta_{b_2 d_2}+ \delta_{a_2 b_2}\delta_{d_2 c_2}) \big) \crcr
		&\quad \, + \, \frac{g_d}{3} \big( \delta_{a_1b_1} \delta_{a_2b_2} \delta_{c_1d_1} \delta_{c_2d_2} \, + \, \delta_{a_1c_1} \delta_{a_2 c_2} \delta_{b_1d_1} \delta_{b_2d_2} \,+\, \delta_{a_1d_1} \delta_{a_2 d_2} \delta_{c_1b_1} \delta_{c_2b_2}\big) \, , \\
		\label{eq:r-bifund}
		r_{\mba \mbb} \, &= \, r \, \delta_{a_1b_1} \delta_{a_2b_2} \, ,
	\end{align}
where each boldface index is split in a pair of indices, $\mba=(a_1, a_2)$ and so on,
where the first index corresponds to the $O(M)$ group ($a_1 = 1, \ldots, M$) while the second index corresponds to the $O(N)$ group ($a_2 = 1, \ldots, N$). This model can be viewed as a modification of the $O(\cN)$ model by the $g_s$ term, which  explicitly breaks the $O(MN)$ symmetry down to $O(M)\times O(N)$.
It can of course also be viewed as a rectangular matrix field theory, with $g_s$ being associated to the single-trace interaction $\Tr[\phi \phi^t \phi \phi^t]$ and $g_d$ to the double-trace interaction $\Tr[\phi \phi^t]^2$, thus explaining our choice of subscripts.

Plugging \eqref{eq:g-bifund} and \eqref{eq:r-bifund}  into \eqref{eq:beta_abcd_alpha} and \eqref{eq:beta2_abcd_alpha}, we obtain the beta functions\footnote{Since the three-loop contributions to the beta functions are too lengthy, we do not display them here. They will however be taken into account for the analysis of the fixed points.}
	\begin{align}
		\beta^{(4)}_s \, &= \, - \epsilon \, \gt_s \, + \, \frac{\alpha_{D}}{3} \Big[ (M+N+4)\gt_s + 12\gt_d \Big] \gt_s \\
		&\quad + \, \frac{\alpha_{S}}{9} \Big[ (2MN+5M+5N+27) \gt_s^2 + 6(MN+ 14) \gt_d^2 + 12 (2M+2N+5) \gt_d\gt_s \Big] \gt_s \, , \nonumber\\
		\beta^{(4)}_d \, &= \, - \epsilon \, \gt_d \, + \, \frac{\alpha_{D}}{3} \Big[ 3\gt_s^2 +2 (M+N+1) \gt_s \gt_d + (MN + 8) \gt_d^2 \Big] \\
		&\quad + \, \frac{\alpha_{S}}{9} \Big[ 3(M+N+3) \gt_s^3 + 3 (MN + M + N + 15) \gt_s^2 \gt_d  \nonumber \\
		&\qquad\quad + 24(M + N + 1) \gt_s \gt_d^2 + 2 (5MN+22) \gt_d^3 \Big] \, , \nonumber\\
		\beta^{(2)} \, &= \, - (d-2\Delta_{\phi} ) \, \rt \, + \, \frac{\alpha_{D}}{3} \, \Big[ (M+N+1) \gt_s + (MN+2) \gt_d \Big] \, \rt \\
		&\quad + \, \frac{\alpha_S}{6} \, \Big[ (MN+M+N+3) \gt_s^2 + 4 (M+N+1) \gt_s \gt_d + 2(MN+2) \gt_d^2 \Big] \, \rt \, . \nonumber
	\end{align}
At order $\epsilon$, the critical couplings are given by
	\begin{align}
		(\tilde{g}_s^{\star}, \tilde{g}_d^{\star}) \, = \, & (0,0) \, , \, \left(0, \, \frac{3\epsilon}{MN+8} \right) \, , \\
		 & \bigg( \frac{(12-3MN) \epsilon}{4+10(M+N)-MN(M+N+4) \pm 6 \sqrt{Q} } \, , \, \nonumber \\
		& \quad   -  \frac{3}{2} \, \frac{(-80+2M+2N+M^2+N^2+2MN\mp 4(M+N+4)\sqrt{Q}) \epsilon}{464-56(M+N)-16(M^2+N^2+MN)+8MN(M+N) +MN(M+N)^2} \bigg) \, ,
		\nonumber
	\end{align}
where
	\begin{align}
		Q \, = \, 52 - 4 (M+N) + (M^2 - 10MN +N^2) \, .
	\end{align} 
The first solution is the trivial one, and the second solution is the Heisenberg  fixed-point with  $O(MN)$ symmetry.
The third and fourth solutions are the chiral and anti-chiral fixed-points \cite{Kawamura:1988}. When $\tilde{g}_s^{\star}<0$, the latter are also called sinusoidal and anti-sinusoidal fixed points.

There are four regimes of criticality at fixed $M$  depending on the stability of the Heisenberg and chiral fixed points:
\begin{itemize}
\item If $N> N_{c+}$ there are four real fixed points, and the chiral one is stable.
\item If $N_{c-}<N< N_{c+}$, only the Gaussian and the Heisenberg fixed points are real, and they are both unstable.
\item If $N_H <N < N_{c-}$, there are again four real fixed points, and the chiral (or sinusoidal) one is stable.
\item If $N<N_H$, there are still four real fixed points, but the Heisenberg one is stable. 
\end{itemize}

To compute $N_{c\pm}$ and $N_H$, we use the following ansatz:
	\begin{align}
		N \, = \, N_0 \, + \, N_1\, \epsilon \,+ \, N_2 \, \epsilon^2 \, + \, \mathcal{O}(\epsilon^3) \, ,
	\end{align} 
	
and we solve:
\begin{align}
		{\rm det} \left| \frac{\partial (\beta_s, \beta_d)}{\partial (\gt_s, \gt_d)} \right|_{\gt=\gt^\star} \, = \, 0 \, .
	\end{align} 

This leads to $N_{c\pm}=N_{c\pm,0}+N_{c\pm,1}\epsilon+N_{c\pm,2}\epsilon^2+\mathcal{O}(\epsilon^3)$ with:
	\begin{align}
		N_{c\pm,0} \, = \, 2 + 5M \pm 2 \sqrt{6(M+2)(M-1)} \, ,
	\end{align} 
	\begin{align}
		N_{c\pm,1}\, = \, (5M+2)\alpha_{S,0}\pm \frac{\alpha_{S,0}}{2s}\Big[25M^2+22M-32\Big] \, ,
	\end{align}

\begin{align}
N_{c\pm,2}\,&= \, \frac{\alpha_{I_1}}{Q_2}\Big[\frac{5M^5+14M^4-277M^3-530M^2+496M+400}{3}\crcr
& \qquad \quad \pm\frac{s(25M^4+59M^3-1434M^2-1900M+2944)}{36}\Big] \crcr
& +\frac{\alpha_{I_2}}{Q_2}\Big[\frac{2(5M^5+11M^4-283M^3-341M^2+418M+136)}{3}\crcr
& \qquad \quad\pm\frac{s(25M^4+41M^2-1398M^2-946M+1216)}{18}\Big]\crcr
&+\frac{\alpha_{I_4}}{Q_2}\Big[\frac{8M^5+31M^4-426M^3-1376M^2+1184M+1632}{96}\crcr
& \qquad \quad\pm\frac{s(20M^4+73M^3-1230M^2-2960M+6176)}{576}\Big] \crcr
& +\frac{\alpha_{T}}{Q_2}\Big[\frac{5M^5+8M^4-289M^3-152M^2+340M-128}{12}\crcr
& \qquad \quad\pm\frac{s(25M^4+23M^3-1362M^2+8M-512)}{144}\Big] \crcr
& +\frac{\alpha_{U}}{Q_2}\Big[\frac{25M^5+64M^4-1397M^3-2272M^2+2324M+1472}{24}\crcr
& \qquad \quad\pm\frac{s(125M^4+259M^3-7098M^2-7592M+11264)}{288}\Big] \crcr
& +\left(\alpha_{S,1}-2\alpha_{S,0}\alpha_{D,1}\right)\Big[5M+2\mp\frac{3(M-6)}{2s}\pm\frac{25s}{12}\Big] \crcr
& -\frac{\alpha_{S,0}^2}{Q_1}\Big[\frac{7M^7+32M^6-744M^5-2882M^4+21608M^3+62520M^2-61952M-62464)}{8} \crcr
& \qquad \quad \pm\frac{1}{96s}\Big(235M^8+1180M^7-26243M^6-108344M^5+791476M^4\crcr
& \qquad \qquad \qquad  +2530384M^3-3402944M^2-6391808M+6897664\Big)\Big] \,,
\end{align}

with
	\begin{align}
s \, &= \, \sqrt{6(M+2)(M-1)} \crcr
Q_1 \, &=\, (M+8)^2(M-7)^2(M+2)(M-1) \crcr
Q_2\, &=\, (M+8)(M-7)(M+2)(M-1) \, .
	\end{align} 
	
For $N_H$, we find:
\begin{align}
N_H=\frac{4}{M}+\frac{2\alpha_{S,0}}{M}\epsilon+\frac{4(8\alpha_{I_1}+4\alpha_{I_2}-\alpha_T)+5\alpha_{I_4}+48(\alpha_{S,1}-2\alpha_{D,1}\alpha_{S,0})-52\alpha_{S,0}^2+14\alpha_{U}}{24M}\epsilon^2+\mathcal{O}(\epsilon^3)\,.
\end{align}

We notice again that the two-loop results for the critical values of $N$ (order $\epsilon$) coincide with the corresponding short-range results \cite{Pelissetto:2001fi} upon taking $\alpha_{S,0}\to -1$.  The three-loop results are instead not related in a similar way.

We can look at the numerical values of $N_H$ and $N_{c\pm}$ at $d=3$ and $M=2$. Table \ref{tab:ncp} 
gives values of $N_{c\pm}$ and $N_H$ for different values of $\epsilon$ with either a three-loop truncation or a Pad\'e-Borel summation method. 
The table indicates that for $d=3$ and $M=N=2$ the chiral (or sinusoidal) fixed point might exist and be stable for sufficiently small $\epsilon$. However, at $N=3$ the chiral fixed point is not present, and the Heisenberg one is not stable.  
\begin{table}[h]
\begin{center}
\begin{tabular}{|c|c||c|c|c|}\hline
$\epsilon$ &  & one-loop &        three-loop   &    PB $[1/1]$ \\ \hline
\multirow{3}{*}{0.2} & $N_{c+}$ & 21.8 &   16.36  &  15.7(12) \\
                     & $N_{c-}$ & 2.202 &  2.120  &  2.076(35) \\
                     & $N_{H}$ &   2  &  1.806  &  1.759(12)  \\
\hline
\multirow{3}{*}{0.4} & $N_{c+}$ & 21.8 &   15.35  &  11.6(29) \\
                     & $N_{c-}$ & 2.202 &  2.245  &   2.01(9) \\
                     & $N_{H}$ &   2  &   1.875      &  1.608(29)  \\
\hline
\multirow{3}{*}{0.6} & $N_{c+}$ & 21.8 &  18.76   &  8(4)     \\
                     & $N_{c-}$ & 2.202 & 2.578 &  1.96(14) \\
                     & $N_{H}$ &   2  &  2.207  &  1.50(5) \\
\hline
\end{tabular}\end{center}
\caption{The critical values $N_{c\pm}$ and  $N_{H}$ for the long-range bifundamental model at $d=3$ and $M=2$, as computed by a one-loop and three-loop truncation and by a Pad\'e-Borel summation of the three-loop series with $[1/1]$ approximant (with error estimated by the difference with the PB summation of the two-loop series with $[0/1]$ approximant).}
\label{tab:ncp}
\end{table}


\section{Conclusions}
\label{sec:concl}

Long-range models have several intriguing properties and they provide an interesting playground for statistical physics methods.
Nevertheless, they have been much less explored than their short-range counterparts. In particular, fewer models have been considered, and all perturbative results to date had been limited to two loops.
In this work we contributed in two ways to improving the situation in the case of long-range multi-scalar models: first, we computed the renormalization group beta functions for general quartic interaction up to three loops; second, we used them to provide higher-order results in the long-range Ising, $O(N)$, cubic, as well as $O(M)\times O(N)$ models.
Along the way, we showed that the hypothetical relations between the long-range Ising model at given dimension $d$ and the short-range Ising model at a different dimension $d_{SR}$ \cite{Banos:2012,Angelini:2014,Defenu:2014} only hold up to first order in $\epsilon_{SR}=4-d_{SR}$, failing at second order.

It is instructive to compare our computations to the analogue three-loop computations for short-range multi-scalar models by Brezin et al.\ \cite{Brezin:1974eb,Brezin:1974-add,Brezin:1973jt}. The setting is very similar to ours, as we do not use the minimal subtraction scheme, relying instead on renormalization conditions at a subtraction point. On the latter we differ, as for the four-point function they adopted a non-zero symmetric subtraction point in momentum space, preserving the massless propagator while avoiding IR divergences in the renormalization condition. Unfortunately, for our integrals this option turned out to be unfeasible: we have of course the same topologies of Feynman diagrams as they do, but with propagators with an essentially arbitrary power of $1/p^2$. While in the short-range case the three-loop integrals can be performed for example in Feynman parametrization, the same representation includes in our case extra factors $u_i^{\z-1}$ 
subject to the constraint $\sum_i u_i=1$: the constraint makes it difficult to perform the explicit integration over the parameters $u_i$. Similarly, the Schwinger parametrization with symmetric subtraction point leads to inconvenient exponents of rational functions.
Therefore, we opted for a subtraction point at zero external momenta, and were forced to introduce an IR regulator in the propagator, equation \eqref{eq:param}. We then used a Mellin-Barnes representation for the Schwinger parametrization of the amplitudes, with the sole exception of the $I_4$ integral of appendix \ref{app:I_4}, for which we used the Gegenbauer polynomial technique \cite{Chetyrkin:1980pr} applied directly to the integral in  momentum space. All these details are presented in the appendices \ref{sec:MB} and \ref{app:integrals}. 
We want to emphasize that despite the vast literature on Feynman integral calculus (e.g.\ \cite{Smirnov:2006ry,Weinzierl:2006qs,Panzer:2014kia,Kotikov:2018wxe} and relative encyclopedic work \cite{Bogner:2017xhp}) we were unable to find directly applicable methods besides the ones presented here. Of course this could be due to our limitations, and it would be interesting to further explore the evaluation of long-range Feynman integral with other methods, possibly allowing a massless renormalization scheme or higher-loop computations.

Moreover, we hope that our higher-order results will provide further motivation for studying  long-range models also by other methods, such as Monte-Carlo simulations (at $d=3$ in particular), bootstrap, or functional RG. Some work by such means on long-range Ising and $O(N)$ models has been initiated (see \cite{Angelini:2014,Defenu:2014,Behan:2018hfx}) but clearly there is room and motivation for more.

We conclude with one comment about the crossover from long-range to short-range critical behavior for more general models than Ising. For the latter, such crossover has attracted quite some interest \cite{Sak:1973,Blanchard:2012xv,Angelini:2014,Brezin:2014,Defenu:2014,Behan:2017dwr,Behan:2017emf}, with the emerging picture being that as one varies $\z$ in the $(0,1)$ interval, at fixed $d$, three regimes are met: the long-range mean-field regime for $0<\z<d/4$; the long-range non-trivial critical regime for $d/4<\z<\z^\star$; and the short-range critical regime for $\z>\z^\star$. The value $\z^\star$ is such that $2\z^\star = 2 -\eta_{\rm SR}$, where $\eta_{\rm SR}$ stands for the anomalous dimension of the short-range Ising model in $d$ dimensions. The picture found in \cite{Behan:2017dwr,Behan:2017emf} actually suggests that the crossover at $\z=\z^\star$ happens not just to the short-range fixed point, but rather to the short-range fixed point plus a decoupled Gaussian field. It would be interesting to extend such picture to the general multi-scalar models.

\section*{Acknowledgements}

The work of DB, RG and SH is supported by the European Research Council (ERC) under the European Union's Horizon 2020 research and innovation program (grant agreement No818066).
The work of KS is supported by the European Research Council (ERC) under the European Union's Horizon 2020 research and innovation program (grant agreement No758759). 

This work was partly supported by Perimeter Institute for Theoretical Physics.

\cleardoublepage

\appendix

\section{The renormalized series} 
\label{app:renseries}

The inversion of the bare series is immediate using the Bogoliubov Parasuk recursion \cite{Rivasseau:1991ub}. The renormalized series is identical to the bare one, up to exchanging the roles of the renormalized and bare constants and replacing the bare amplitudes by counterterms:
\be
\begin{split}
 g_{\mba \mbb\mbc\mbd} 
 & = \mu^{-\epsilon} \lambda_{\mba \mbb\mbc\mbd} - \sum_G
 s(G)( - 1)^V \mathbf{G}( \mu^{-\epsilon}\lambda)_{\mba\mbb\mbc\mbd} \; {\cal \hat A}(G) \,, \crcr
  \mu^{-\epsilon} \lambda_{\mba \mbb\mbc\mbd} 
 & = g_{\mba \mbb\mbc\mbd} - \sum_G
 s(G)(-1)^V \mathbf{G}( g )_{\mba\mbb\mbc\mbd} \; K_G \,,
\end{split}
\ee
where the sums run over one particle irreducible four-point graphs $G$ and
\begin{itemize}
 \item  $V$ is the number of vertices of $G$ and 
 $s(G)$ is the symmetry factor of $G$,
 \item $ \mathbf{G}( \cdot )_{\mba\mbb\mbc\mbd} $ is the contraction of coupling constants associated to $G$, which in our case has four external points with four associated external indices $\mba\mbb\mbc\mbd$,
 \item $ {\cal \hat A}(G)$ is the bare amplitude in \eqref{eq:amp_final} and $K_G$ is the counterterm of $G$ which we define below.
\end{itemize}

The point is that $s(G)$ and $\mathbf{G}(\cdot)_{\mba\mbb\mbc\mbd}$ are the \emph{same} in the two formulae.
The counterterms $K_G$ are defined recursively:
\be
\qquad K_G = - \sum_{G_1,\dots G_k} 
   {\cal \hat A} ( G/{\cup G_i} )  \prod_{i} K_{G_i} \,,
\ee
where the sum runs over all the families of (vertex) disjoint one-particle irreducible four-point subgraphs\footnote{In general one sums over all primitively divergent subgraphs, but as we subtracted the two-point graphs to zero this reduces to the 1PI four-point subgraphs.} $G_i$ of $G$, including the empty family ($G$ itself is not its own subgraph), and $ G/\cup G_i$ is the graph $G$ where all the $G_i$ have been contracted to four-point vertices. This definition is recursive: the counterterms $K_{G_i}$ have already been defined at this stage.

At one loop $D$ has no four-point subgraphs, hence 
$ K_D = -D$. At two loops, $D^2$ has two subgraphs $D$ (sharing a vertex), while $S$ has one subgraph $D$, hence:
\be
K_{D^2} = - D^2  - 2 D K_D = D^2 
 \,,\qquad 
 K_S = - S - D  K_D = D^2 - S
 \,,
\ee
and a short computation yields at three loops:
\be
\begin{split}
& K_{D^3} = -D^3 \,, \quad
K_{DS} = -D (D^2-S) \,, \quad
K_{U} = -U + 2SD - D^3 \,,\quad
 K_{T} = -T + 2DS - D^3\,, \crcr
& K_{I_1}  = -I_1 + 2SD -D^3 \,, \quad 
K_{I_2} = -I_2 + 2DS -D^3 \,, 
\quad K_{I_3} = -I_3 \,,\quad K_{I_4} = -I_4 \,.
 \end{split}
\ee
Observe that $I_3$ has four-point subgraphs, but upon contracting we obtain a graph with a tadpole hence with zero bare amplitude. Therefore, the inverse of \eqref{eq:bphz}, with right-hand sides given in \eqref{eq:bare_4pt} and \eqref{eq:bare_2pt}, are:
\begin{align}
& \mu^{-\epsilon}\lambda_{\mba \mbb \mbc \mbd}
\, = \, g_{\mba \mbb \mbc \mbd}+\frac{1}{2} \big(g_{\mba \mbb \mbe \mbf}g_{\mbe \mbf \mbc \mbd} + 2 \textrm{ terms} \big) \, D \crcr
& \quad
+ \, \frac{1}{4} \big(g_{\mba \mbb \mbe \mbf} g_{\mbe \mbf \mbg \mbh} g_{\mbg \mbh \mbc \mbd} + 2 \textrm{ terms} \big) \,D^2 \, 
+ \,\frac{1}{2} \big(g_{\mba \mbb \mbe \mbf} g_{\mbe \mbg \mbh \mbc} g_{\mbf \mbg \mbh \mbd} + 5 \textrm{ terms} \big)  \, (D^2 - S)   \crcr
&\quad
+ \, \frac{1}{8} \big(g_{\mba \mbb \mbe \mbf} g_{\mbe \mbf \mbg \mbh} g_{\mbg \mbh \mbm \mbn} g_{\mbm \mbn \mbc \mbd} 
+ 2 \textrm{ terms} \big) \, D^3  \, 
+ \, \frac{1}{4} \big(g_{\mba \mbb \mbe \mbf} g_{\mbe \mbf \mbg \mbh} g_{\mbg \mbm \mbn \mbc} g_{\mbh \mbm \mbn \mbd} 
+ 5 \textrm{ terms} \big) \, D(D^2-S)   \crcr
& \quad 
+ \, \frac{1}{4} \big(g_{\mba \mbe \mbf \mbg} g_{\mbb \mbe \mbf \mbh} g_{\mbg \mbm \mbn \mbc} g_{\mbh \mbm \mbn \mbd} + 5 \textrm{ terms} \big) \; (D^3-2DS+ U)   \crcr
& \quad   + \, \frac{1}{4} \big(g_{\mba \mbb \mbe \mbf} g_{\mbe \mbg \mbh \mbm} g_{\mbf \mbg \mbh \mbn} g_{\mbm \mbn \mbc \mbd} + 2 \textrm{ terms} \big) (D^3-2DS+ T)  \crcr
& \quad   + \,\frac{1}{2} \big(g_{\mba \mbb \mbe \mbf} g_{\mbe \mbg \mbh \mbm} g_{\mbf \mbg \mbn \mbc} g_{\mbh \mbm \mbn \mbd}  + 11 \textrm{ terms} \big) \, (D^3-2DS+I_1)   \crcr
& \quad
+ \,  \frac{1}{4} \big(g_{\mba \mbb \mbe \mbf} g_{\mbe \mbg \mbh \mbc} g_{\mbf \mbm \mbn \mbd} g_{\mbg \mbh \mbm \mbn} 
+ 5 \textrm{ terms} \big) (D^3-2DS+ I_2)    \crcr
& \quad  + \, \frac{1}{6} \big(g_{\mba \mbb \mbe \mbf}g_{\mbh \mbm \mbn \mbf}g_{\mbh \mbm \mbn \mbg}g_{\mbg \mbe \mbc \mbd} +2 \textrm{ terms} \big) \, I_3  + \,  \big( g_{\mba \mbe \mbm \mbh}g_{\mbb \mbe \mbf \mbn}g_{\mbc \mbf \mbm \mbg}g_{\mbd \mbg \mbn \mbh}  \big) \, I_4\, ,
\label{eq:inverse}
\end{align}
while for the two-point coupling we get:
\begin{equation}
\begin{split}
& \mu^{-(d-2\Delta_{\phi})}\kappa_{\mbc \mbd} \, =  \, 
r_{ \mbc \mbd} + \frac{1}{2} \big( r_{\mbe \mbf} g_{\mbe \mbf \mbc \mbd} \big) \, D 
+\frac{1}{4} \big( r_{\mbe \mbf} g_{\mbe \mbf \mbg \mbh}g_{\mbg \mbh \mbc \mbd} \big) \, D^2 
+\frac{1}{2} \big( r_{\mbe \mbf} g_{\mbe \mbg \mbh \mbc}g_{\mbf \mbg \mbh \mbd} \big) \, (D^2-S) \crcr 
&
+ \, \frac{1}{8}  \big( r_{ \mbe \mbf} g_{\mbe \mbf \mbg \mbh} g_{\mbg \mbh \mbm \mbn} g_{\mbm \mbn \mbc \mbd} \big) \, D^3  
+ \, \frac{1}{4} \big( r_{ \mbe \mbf} g_{\mbe \mbf \mbg \mbh} g_{\mbg \mbm \mbn \mbc} g_{\mbh \mbm \mbn \mbd} \big) \,  
D(D^2-S) \crcr 
& + \, \frac{1}{4} \big( r_{ \mbe \mbf} g_{\mbe \mbg \mbh \mbm} g_{\mbf \mbg \mbh \mbn} g_{\mbm \mbn \mbc \mbd} \big) \, 
( D^3 -2DS  + T ) + \, \frac{1}{2} \big( r_{ \mbe \mbf} g_{\mbe \mbg \mbh \mbm} g_{\mbf \mbg \mbn \mbc} g_{\mbh \mbm \mbn \mbd}  + 1 \textrm{ term} \big) \,(D^3 - 2DS + I_1 )\crcr 
&
+ \,  \frac{1}{4} \big( r_{ \mbe \mbf} g_{\mbe \mbg \mbh \mbc} g_{\mbf \mbm \mbn \mbd} g_{\mbg \mbh \mbm \mbn} \big) \, (D^3 - 2DS +  I_2 ) 
+ \, \frac{1}{6} \big(r_{ \mbe \mbf}\lambda_{\mbh \mbm \mbn \mbf}\lambda_{\mbh \mbm \mbn \mbg}\lambda_{\mbg \mbe \mbc \mbd}   \big)\,   I_3  \, .
\end{split}
\end{equation}

In order to compute the $\beta$ functions in practice,
we can for instance derive the bare series in \eqref{eq:bare_4pt} and \eqref{eq:bare_2pt} with respect to $\mu$ and then substitute the bare constants in terms of the renormalized ones using the renormalized series. We get:
\begin{align}
\beta^{(4)}_{\mba \mbb \mbc \mbd} 
&= -\epsilon g_{\mba \mbb \mbc \mbd} +\frac{\epsilon}{2}  D\, \big(g_{\mba \mbb \mbe \mbf}g_{\mbe \mbf \mbc \mbd} + 2 \textrm{ terms} \big)  + 
\frac{\epsilon}{2} \left(D^2-2S\right) \,
\big(g_{\mba \mbb \mbe \mbf}g_{\mbe \mbg \mbh \mbc}g_{\mbf \mbg \mbh \mbd}+ 5 \textrm{ terms}\big) \crcr 
& \quad  \, + \, \frac{\epsilon}{4} (D^3-4DS+3U) \,
\big( g_{\mba \mbe \mbf \mbg} g_{\mbb \mbe \mbf \mbh} g_{\mbg \mbm \mbn \mbc} g_{\mbh \mbm \mbn \mbd} + 5 \textrm{ terms} \big)  \crcr
& \quad  \, + \, \frac{\epsilon}{4}(3T-2DS) \, \big(g_{\mba \mbb \mbe \mbf} g_{\mbe \mbg \mbh \mbm} g_{\mbf \mbg \mbh \mbn} g_{\mbm \mbn \mbc \mbd} + 2 \textrm{ terms} \big)  \crcr
& \quad  \, + \, \frac{\epsilon}{2}(D^3-3DS+3I_1) \, \big(g_{\mba \mbb \mbe \mbf} g_{\mbe \mbg \mbh \mbm} g_{\mbf \mbg \mbn \mbc} g_{\mbh \mbm \mbn \mbd} + 11 \textrm{ terms} \big)  \crcr
& \quad  \, + \, \frac{\epsilon}{4}(D^3-4DS+3I_2) \,
\big(g_{\mba \mbb \mbe \mbf} g_{\mbe \mbg \mbh \mbc} g_{\mbf \mbm \mbn \mbd} g_{\mbg \mbh \mbm \mbn} + 5 \textrm{ terms} \big)  \crcr
& \quad  + \, \frac{\epsilon}{2} I_3 \, \big(g_{\mba \mbb \mbe \mbf}g_{\mbh \mbm \mbn \mbf}g_{\mbh \mbm \mbn \mbg}g_{\mbg \mbe \mbc \mbd} +2 \textrm{ terms} \big) \,  + \, 3\epsilon I_4 \,\big( g_{\mba \mbe \mbm \mbh}g_{\mbb \mbe \mbf \mbn}g_{\mbc \mbf \mbm \mbg}g_{\mbd \mbg \mbn \mbh}  \big) \,,
\end{align}
while for the quadratic coupling we get
\begin{align}
\beta^{(2)}_{\mbc \mbd} 
&= - (d-2\Delta_{\phi} ) r_{\mbc \mbd} +\frac{\epsilon}{2}  D\big( r_{ \mbe \mbf}g_{\mbe \mbf \mbc \mbd} \big) + \frac{\epsilon}{2} \left(D^2-2S\right)
 \big(r_{\mbe \mbf}g_{\mbe \mbg \mbh \mbc}g_{\mbf \mbg \mbh \mbd} \big) \crcr 
& \quad  \,  + \, \frac{\epsilon}{4}(3T-2DS) 
\big(r_{\mbe \mbf} g_{\mbe \mbg \mbh \mbm} g_{\mbf \mbg \mbh \mbn} g_{\mbm \mbn \mbc \mbd} \big) + \, \frac{\epsilon}{2}(D^3-3DS+3I_1) 
(r_{\mbe \mbf} g_{\mbe \mbg \mbh \mbm} g_{\mbf \mbg \mbn \mbc} g_{\mbh \mbm \mbn \mbd} + 1 \textrm{ term} )  \crcr
& \quad  \,  + \, \frac{\epsilon}{4}(D^3-4DS+3I_2) 
\big(r_{ \mbe \mbf} g_{\mbe \mbg \mbh \mbc} g_{\mbf \mbm \mbn \mbd} g_{\mbg \mbh \mbm \mbn} \big) 
 + \, \frac{\epsilon}{2} I_3 
\big( r_{\mbe \mbf} g_{\mbh \mbm \mbn \mbf}g_{\mbh \mbm \mbn \mbg}g_{\mbg \mbe \mbc \mbd} \big) \,.
\label{eq:gamma_ab}
\end{align}

\section{Mellin-Barnes representation}
\label{sec:MB}

We briefly review the Mellin-Barnes representation used repeatedly in appendix \ref{app:integrals}. The Mellin transform of a function $f$ and the inverse Mellin transform are:
 \be
  \phi(s) = \int_0^{\infty} dx  \; x^{s-1} f(x)  \,, \qquad
  f(x) = \int_{c-\imath \infty}^{c+\imath \infty} \frac{ds}{2\pi \imath}\; x^{-s}\phi(s) \,,
 \ee
with $c$ such that $\phi(s)$ is analytic and decreases at infinity in a strip around $c$. In particular, changing variables in the inverse Mellin transform we have:
\be
 \Gamma(s) = \int_0^{\infty} dx\; x^{s-1} \, e^{-x} \,, \qquad
  e^{-x} = \int_{0^--\imath \infty}^{0^-+\imath \infty}
    \frac{ds}{2\pi \imath} \; x^{s} \, \Gamma(-s) \,,
\ee
that is, the $\Gamma$ function is the Mellin transform of the exponential.\footnote{The first expression is the definition of the $\Gamma$ function while the second one is obtained by going around the poles of $\Gamma(-s)$ located at $s=n$, taking into account that the contours are negatively oriented and that 
$
 {\rm Res} \big( \Gamma(-s) ,n \big) = \lim_{s\to  n} (s - n) \Gamma(- s ) = - \lim_{z\to -n} (z + n) \Gamma(z) = - (-1)^n / n!$.} 
 For ${\rm Re}(u) > 0 $ we have:
\begin{equation}
   \frac{ \Gamma(u) }{(A+B)^u}  = \int_{0}^{\infty} dx \int_{0^- -\imath \infty}^{0^- +\imath \infty}
    \frac{dz }{2\pi \imath}
    \;x^{u-1} (x A)^{z} \Gamma(-z ) e^{-xB } 
     =\int_{0^--\imath \infty}^{0^-+\imath \infty} \; 
    \frac{dz }{2\pi \imath} \; \Gamma(-z)  \Gamma(u+z) A^z  B^{-u-z} \,.
\end{equation}
Denoting $[dz] = \frac{dz}{2\pi \im}$ we get:
 \be
 \begin{split}
 &  \frac{1}{(A_1 + \dots + A_{q+1})^{u}}  = \crcr 
 & \qquad \int_{0^- -\imath \infty}^{0^- +\imath \infty} [dz] \;
   \frac{\Gamma(-z_1) \dots \Gamma(-z_q)  \Gamma(z_1 + \dots +z_q +u)  }{\Gamma(u)}  \;  A_1^{z_1} \dots A_q^{z_q} 
   A_{q+1}^{-z_1 - \dots -z_q -u} \,,
 \end{split}
 \ee
which is the main formula we use in appendix \ref{app:integrals}. The only a priori restriction we have on the Mellin contour is that ${\rm Re} (z_i) <0$.

\paragraph{A general remark.} 
Let us assume we are interested in the $\epsilon \to 0$ limit of an integral of the type:
\[
 \int_{C}[dz] \; \frac{H_{\epsilon}(z)}{(z-z_0)(z-z_0-\epsilon)} \,,
\]
where $H_{\epsilon}(z)$ depends parametrically on $\epsilon$ and the two poles (of order one) at $z_0$ and $z_0+\epsilon$ are the only poles inside the contour $C$. In the $\epsilon \to 0$ limit the two poles collapse. The integral has a well-defined $\epsilon \to 0$ limit and can be computed by the residue at the double pole if the function is analytic around $z_0$ uniformly in $\epsilon$ (that is the Taylor series of the function around $z_0$ has a radius of convergence which does not depend on $\epsilon$). To see this, observe that:
\[
 \int_{C}[dz] \; \frac{H_{\epsilon}(z)}{(z-z_0)(z-z_0-\epsilon)}
 = H_{\epsilon}(z_0) \frac{1}{-\epsilon} + H_{\epsilon}(z_0+\epsilon) \frac{1}{\epsilon} = H_{\epsilon}'(z_0) + O(\epsilon) = 
 \lim_{z\to z_0} \frac{d}{dz}\left[ (z-z_0)^2 H_0(z)\right] + O(\epsilon)\,.
\]
where we recall that the residue at a pole of order $n$ is ${\rm Res}(f,c) = 
\frac{1}{(n-1)!} \lim_{z\to c} [(z-c)^n f]^{(n-1)} $. 
The $\epsilon \to 0$ limit does not exist if 
$H_\epsilon(z_0)$ or its derivative diverges in the $\epsilon\to 0$ limit.

\section{The integrals}
\label{app:integrals}

In this appendix we compute the integrals appearing in the three loops beta function.

\subsection{One loop integral $D$}

First, let us compute the amplitude $D$ of the one loop graph. From \eqref{eq:amp_final} this is:
\begin{equation}
D=\frac{1}{(4\pi)^{d/2}\Gamma(\zeta)^2}\int_{0}^{\infty} 
da_1da_2 \; \frac{(a_1 a_2)^{\zeta-1} e^{-\sum a}}{ (a_1 + a_2)^{d/2}}  \,.
\end{equation}
We will repeatedly use below the integral:
\be\label{eq:I} 
  \int_{0}^{\infty}[da] \;  
 \frac{ (a_1 a_2)^{ u -1}}{(a_1+a_2)^{ \gamma }} \;
 e^{- (a_1+a_2) } =  \frac{\Gamma(u)^2 \Gamma(2u-\gamma)}{\Gamma(2u)} \,,
\ee
which is convergent for $2{\rm Re}( u ) > {\rm Re}(\gamma)$ and ${\rm Re}(u)>0$. $D$ is the particular case $u = \zeta, \gamma = d/2$:
\begin{align}
\label{eq:D_tot}
\boxed{
  D  =  \frac{ 1 }{(4\pi)^{d/2}\Gamma(2\zeta)} \Gamma(\tfrac{\epsilon}{2} )  \,. }
\end{align}
At the relevant orders in $\epsilon$ this is: 
\begin{align}
D&=\frac{1}{\left(4\pi\right)^{d/2}\Gamma(\tfrac{d}{2})}\left(\frac{2}{\epsilon}-\psi(\tfrac{d}{2})+\psi(1) +\frac{\epsilon}{24}(6(\psi(1)-\psi(\tfrac{d}{2}))^2+\pi^2-6\psi_1(\tfrac{d}{2}))\right) +\mathcal{O}(\epsilon^2) \,,\crcr
 \alpha_{D}& = \epsilon(4\pi)^{d/2}\Gamma(\tfrac{d}{2}) \; \frac{D}{2}\crcr
  &= 1 +\frac{\epsilon}{2}\left(\psi(1)-\psi(\tfrac{d}{2})\right)+\frac{\epsilon^2}{8}\left(\left(\psi(1)-\psi(\tfrac{d}{2})\right)^2+\psi_1(1)-\psi_1(\tfrac{d}{2})\right)
\,,
\label{eq:D_exp}
\end{align}
where we used $\psi_1(1)=\pi^2/6$.

\subsection{Two-loop integral $S$}

For the $S$ graph in Fig.~\ref{fig:1_2_loops}, \eqref{eq:amp_final} yields:
\begin{equation}
S=\frac{1}{(4\pi)^d\Gamma(\zeta)^4}\int_{0}^{\infty} [dadb] \; \frac{(a_1 a_2 b_1 b_2)^{\zeta-1}}{ \big[(a_1 + a_2)(b_1 + b_2 ) + b_1 b_2 \big] ^{d/2 }} \;e^{-(a_1+a_2+b_1+b_2)} \,.
\end{equation}
Using Mellin parameters (see appendix \ref{sec:MB}) we  write:
\be
 \frac{1}{\big[(a_1 + a_2)(b_1 + b_2 ) + b_1 b_2 \big]^{d/2}}
  =  \frac{1}{\Gamma( \tfrac{d}{2}) } \int_{0^-}[dz]  \Gamma(-z)
  \Gamma(\tfrac{d}{2}+z)  
   \;  \frac{ (b_1b_2)^z }{(b_1+b_2)^{\tfrac{d}{2} + z} } \;\frac{1}{(a_1+a_2)^{ \tfrac{d}{2} + z }} \,,
\ee
and we integrate $a$ and $b$ using \eqref{eq:I} to obtain:
\be
 S = \frac{1}{(4\pi)^d\Gamma(2\zeta)
 \Gamma(\tfrac{d}{2})\Gamma(\zeta)^2 } 
\int_{0^-}[dz] \;  \Gamma(-z) \Gamma( \tfrac{d}{2}+z)  
  \frac{\Gamma(z+\zeta)^2}{ \Gamma(2z+2\zeta ) }
\Gamma( \tfrac{\epsilon}{2} + z)  \; 
   \Gamma(\tfrac{\epsilon}{2} -z) \,.
\ee

In the right half complex plane the integrand has poles at $z = n$ and $ n +\epsilon / 2$, for $n\in \mathbb{N}_0$. Passing to the right of the first two poles we get:
\be\label{eq:S1full}
\boxed{
\begin{split}
  S = & \, D^2 + 
  \; \frac{ 1 }{(4\pi)^d \Gamma(2\zeta) \Gamma(\zeta)^2\Gamma( \tfrac{d}{2})  }  \Gamma(\tfrac{d}{2} + \tfrac{ \epsilon}{2})\frac{\Gamma( \tfrac{\epsilon}{2}+\zeta)^2}{ \Gamma(\epsilon+2\zeta ) } \Gamma(-\tfrac{ \epsilon}{2} ) \Gamma(\epsilon) + \frac{J_{\epsilon}(\zeta)}{(4\pi)^d \Gamma(\frac{d}{2})^2 } \,, \crcr
  J_{\epsilon}(\zeta)  = & \, \frac{\Gamma(\tfrac{d}{2})}{\Gamma(\zeta)^2\Gamma(2\zeta) } 
\int_{1^-}[dz] \;  \Gamma(-z) \Gamma(\tfrac{d}{2}+z)  
  \frac{\Gamma(z+\zeta)^2}{ \Gamma(2z+2\zeta ) }
\Gamma( \tfrac{\epsilon}{2} + z)  \; 
   \Gamma(\tfrac{\epsilon}{2} -z) \,. 
\end{split} }
\ee
The crucial point is that $J_{\epsilon}(\zeta)$ has a finite limit for $\epsilon \to 0$:
\be
 J_0( \tfrac{d}{4}) = \frac{1}{\Gamma( \tfrac{d}{4})^2} 
\int_{1^-}[dz] \;  \Gamma( \tfrac{d}{2}+z)  
  \frac{\Gamma(z+ \tfrac{d}{4})^2}{ \Gamma(2z+ \tfrac{d}{2} ) }
\Gamma(  z)  \; 
   \Gamma( -z)^2 \,,
\ee
the integrand having poles of order $2$ with finite residues at the positive integers. $J_0(\tfrac{d}{4})$ can also be expressed as an infinite sum, which we will use for numerical estimates:
\begin{equation}
J_0(\tfrac{d}{4})=\frac{1}{\Gamma(\tfrac{d}{4})^2 }\sum_{n \geq 1}\frac{\Gamma(n+ \tfrac{d}{2})\Gamma(n+ \tfrac{d}{4})^2}{n(n!)\Gamma( \tfrac{d}{2}+2n)} \Big(2\psi(n+1)-\psi(n)-2\psi(n+\tfrac{d}{4})-\psi(n+\tfrac{d}{2})+2\psi(\tfrac{d}{2}+2n)\Big) \,.
\end{equation}
At the relevant order, we obtain for $S$:
\begin{align}
S=& \frac{1}{\left(4\pi\right)^{d}\Gamma(\tfrac{d}{2})^2}\left(\frac{2}{\epsilon^2}+\frac{1}{\epsilon}\Big[3\psi(1)-\psi( \tfrac{d}{2})-2\psi(\tfrac{d}{4})\Big] +\frac{7}{4}\psi(1)^2-\frac{\pi^2}{24}-\psi(1)\psi(\tfrac{d}{4})-\psi(\tfrac{d}{4})^2 \right. \crcr
& \; \left. -\frac{5}{2}\psi(1)\psi(\tfrac{d}{2})+3\psi(\tfrac{d}{4})\psi(\tfrac{d}{2})-\frac{\psi(\tfrac{d}{2})^2}{4}-\psi_1(\tfrac{d}{4})+\frac{5}{4}\psi_1(\tfrac{d}{2})\right)+ \frac{J_0( \tfrac{d}{4})}{(4\pi)^d \Gamma(\tfrac{d}{2})^2} +\mathcal{O}(\epsilon)
\label{eq:S_exp} \,.
\end{align}
We are interested in $\alpha_{S}  = \epsilon (4\pi)^d \Gamma(\tfrac{d}{2})^2\frac{(D^2-2S)}{2}$ for which we get:
\begin{align}
\alpha_{S}  = & \,  2\psi( \tfrac{d}{4} ) - \psi( \tfrac{d}{2})-\psi(1)\crcr
& \, +\frac{\epsilon}{4}\Big[\left(2\psi(\tfrac{d}{4})-\psi(\tfrac{d}{2})-\psi(1)\right)\left(3\psi(1)-5\psi(\tfrac{d}{2})+2\psi(\tfrac{d}{4})\right)  \crcr
& \qquad  +3\psi_1(1)+4\psi_1(\tfrac{d}{4})-7\psi_1(\tfrac{d}{2})-4J_0(\tfrac{d}{4})\Big]   
\,.
\label{eq:alpha_2}
\end{align}

\subsection{Three-loop integrals}

We will treat $I_4$ separately at the end. Using \eqref{eq:amp_final}, the integrals for the other three loops graphs of 
Fig.~\ref{fig:3_loops} are:
\begin{align}\label{eq:SDTU}
 T = & \frac{1}{(4\pi)^{3d/2}\Gamma(\zeta)^6}\int_{0}^{ \infty }  \frac{  (a_1 a_2 b_1 b_2 c_1  c_2)^{\zeta-1}  e^{-\sum a}  }
  { \big[ (a_1 + a_2)(b_1 + b_2) (c_1 + c_2) +  b_1b_2(a_1+a_2+c_1+c_2) \big]^{d/2}  }\,, \crcr
 U = & \frac{1}{(4\pi)^{3d/2}\Gamma(\zeta)^6} \int_{0}^{\infty}   \; \frac{  (a_1 a_2 b_1 b_2 c_1  c_2)^{\zeta-1}  e^{-\sum a}  }
  { \big[ (a_1 + a_2)(b_1 + b_2) (c_1 + c_2) + (a_1 + a_2) c_1c_2 + a_1a_2(c_1+c_2)\big]^{d/2}  } \,, \crcr
I_1 = & \frac{1}{(4\pi)^{3d/2}\Gamma(\zeta)^6}\int_0^{\infty} 
 \frac{ (a_1a_2 b c_1 c_{2'}c_{2''})^{\zeta-1} e^{-\sum a}}
 { \big[ (a_1+a_2)[ c_1( c_{2'} +c_{2''} ) +c_{2'}c_{2''} ] + b [ c_{2'}c_{2''} + ( c_{2'} + c_{2''}) ( a_1 + a_2 + c_1 ) ] \big]^{d/2}} \,,
\crcr
I_2  = & \frac{1}{(4\pi)^{3d/2}\Gamma(\zeta)^6}\int_0^{\infty} 
 \frac{  (a_1a_2b_{1'} b_{1''} b_{2'} b_{2''})^{\zeta-1} e^{-\sum a} }{\big[ (a_1+a_2) ( b_{1'} + b_{1''} )(b_{2'} + b_{2''}  )  + b_{1'} b_{1''} (b_{2'} + b_{2''}) + b_{2'} b_{2''} (b_{1'} + b_{1''}) \big]^{d/2} } \, , \crcr
I_3 = & \frac{1}{(4\pi)^{3d/2}\Gamma(\zeta)^6}\int_0^{\infty} 
 \frac{  ( a_1a_2a_3 b_1 b_2 b_3 )^{\zeta-1} e^{-\sum a} }{\big[ (a_1+a_2+a_3) ( b_1 b_2 + b_1b_3 + b_2b_3) + b_1 b_2 b_3  \big]^{d/2} } \, , 
 \end{align}
where $\sum a$ is an abusive notation which signifies the sum over all the Schwinger parameters.

Notice that $U=I_2$. We thus have only five more integrals to compute. To simplify the notation below, we will call $\tilde{D}=(4\pi)^{d/2}D$, $\tilde{S}=(4\pi)^d S$, $\tilde{T} =(4\pi)^{3d/2} T$ and so on.


\subsubsection{The $T$ integral}
We start with the $T$ integral. We split the denominator using Mellin parameters:
\begin{align}
&\frac{1}
 { [(a_1+a_2)(b_1+b_2)(c_1 + c_2)
+ b_1b_2(a_1+a_2+c_1+c_2) ]^{d/2} }  \crcr
& \qquad = \int_{0^-}[dz_1 ]  \int_{0^-}[ dz_2]
 \frac{\Gamma(-z_1) \Gamma(-z_2) \Gamma( \tfrac{d}{2} +z_1+z_2)}{\Gamma(\tfrac{d}{2})} 
 \frac{ [b_1b_2(a_1+a_2)]^{z_1} [b_1b_2(c_1+c_2)]^{z_2} }
{ [(a_1+a_2)(b_1+b_2)(c_1 + c_2)]^{\tfrac{d}{2} + z_1+z_2} } \,,
\end{align}
and integrating out $a,b,c$ we get:
\begin{align}
 \tilde{T}= \frac{1}{\Gamma(\zeta)^2\Gamma(2\zeta)^2 \Gamma(\tfrac{d}{2})} & \int_{0^-}[dz_1 ]  \int_{0^-}[ dz_2]  \;
  \Gamma(\tfrac{d}{2} +z_1+z_2) 
\;\; \frac{ \Gamma(\zeta + z_1+z_2)^2 } 
{\Gamma(2\zeta + 2z_1+2z_2)}  \Gamma( \tfrac{\epsilon}{2} + z_1 +z_2) \crcr
 & \qquad \qquad \times \Gamma(-z_1) \Gamma(-z_2)   \Gamma( \tfrac{\epsilon}{2}-z_2)    \Gamma( \tfrac{\epsilon}{2}-z_1)
 \,.
\end{align} 
We deform both contours to the right. The poles in $z_1$ and $z_2$ are completely independent, that is for any $z_1$ to the right of $0^-$, the poles in $z_2$ are always located at the values  $n_2$, $n_2+\epsilon/2$. 
We can then push first the contour of $z_2$, pick up the poles in $z_2$ at $z_1$ fixed and then push the contour of $z_1$. Only the poles at $(0,0),(0,\epsilon/2), (\epsilon/2,0)$ and $(\epsilon/2,\epsilon/2)$ give singular contributions. We then obtain: 
\begin{align}
  \tilde{T}  = &
 \frac{1 }{\Gamma(2\zeta)^3 } \Gamma( \tfrac{\epsilon}{2})^3 
 + 2\frac{ \Gamma( \tfrac{d}{2} + \tfrac{\epsilon}{2})\Gamma(\zeta + \tfrac{\epsilon}{2})^2 }{ \Gamma(\zeta)^2\Gamma(2\zeta)^2 \Gamma( \tfrac{d}{2} ) 
 \Gamma(2\zeta +\epsilon) }  \Gamma(\epsilon) \Gamma(-\tfrac{\epsilon}{2}) \Gamma(\tfrac{\epsilon}{2}) \crcr
 & \qquad \qquad + \frac{\Gamma( \tfrac{d}{2} +\epsilon)}{ \Gamma(\zeta)^2\Gamma(2\zeta)^2 \Gamma( \tfrac{d}{2} )}  
 \,  \frac{\Gamma(\zeta +\epsilon)^2}{\Gamma(2\zeta +2\epsilon)} \Gamma( \tfrac{3\epsilon}{2})\Gamma(-\tfrac{\epsilon}{2})^2 
 \crcr
 &+  \frac{2}{\Gamma(\zeta)^2\Gamma(2\zeta)^2 \Gamma(\tfrac{d}{2})}  \int_{1^-}[dz_1 ] 
 \;  \Gamma(-z_1)  \Gamma(\tfrac{\epsilon}{2}-z_1)
  \bigg[
  \Gamma( \tfrac{d}{2}+ z_1) 
\;\; \frac{ \Gamma(\zeta + z_1)^2 } 
{\Gamma(2\zeta + 2z_1)}   \Gamma( \tfrac{\epsilon}{2}+z_1)  \Gamma( \tfrac{\epsilon}{2} )\crcr
 & \qquad \qquad \qquad \qquad \qquad  +  
 \Gamma( \tfrac{d}{2}+ \tfrac{\epsilon}{2} +z_1) 
\;\; \frac{ \Gamma(\zeta + \tfrac{\epsilon}{2} + z_1 )^2 } 
{\Gamma(2\zeta +\epsilon + 2z_1)} 
\Gamma(\epsilon+z_1) \Gamma( - \tfrac{ \epsilon}{2} ) \bigg] 
 \crcr
 &+ \frac{1}{\Gamma(\zeta)^2\Gamma(2\zeta)^2 \Gamma(\tfrac{d}{2})}   \int_{1^-}[dz_1 ]  \int_{1^-}[ dz_2]  \; \Gamma( \tfrac{d}{2} +z_1+z_2) 
\;\; \frac{ \Gamma(\zeta + z_1+z_2)^2 } 
{\Gamma(2\zeta + 2z_1+2z_2)}  \Gamma( \tfrac{\epsilon}{2} + z_1 +z_2) \crcr
 & \qquad \qquad  \qquad \qquad \qquad \times \Gamma(-z_1) \Gamma(-z_2)   \Gamma( \tfrac{\epsilon}{2}-z_2)    \Gamma( \tfrac{\epsilon}{2}-z_1) \,.
 \end{align}
The first two lines are explicit and the single and double integrals are of order $O(\epsilon^0)$.
Overall we get: 
\be
\begin{split}
 T\, &= \, D^3 
 +\frac{\Gamma(-\tfrac{\epsilon}{2})}{(4\pi)^{3d/2}\Gamma(\zeta)^2\Gamma(2\zeta)^2\Gamma( \tfrac{d}{2})} \Bigg[ 
 \frac{2\Gamma( \tfrac{d}{2} +  \tfrac{\epsilon}{2})\Gamma(\zeta + \tfrac{\epsilon}{2} )^2 \Gamma(\epsilon)  \Gamma( \tfrac{\epsilon}{2} )}{\Gamma(2\zeta +\epsilon) } \crcr
&\hspace{120pt} \, + \, \frac{\Gamma( \tfrac{d}{2} + \epsilon)\Gamma(\zeta +\epsilon)^2\Gamma(\tfrac{3\epsilon}{2} )\Gamma(- \tfrac{\epsilon}{2} )}{\Gamma(2\zeta +2\epsilon)} \Bigg] +O(\epsilon^0) \,.
\end{split}
\label{eq:T1_exp}
\ee

At the relevant order in $\epsilon$, this is:

\begin{equation}
\boxed{
\begin{split}
T=\frac{1}{3(4\pi)^{3d/2}\Gamma(d/2)^3}&\left[\frac{8}{\epsilon^3}+\frac{8}{\epsilon^2}\left(2\psi(1)-\psi(\tfrac{d}{4})-\psi(\tfrac{d}{2})\right)\right. \crcr
& \left. +\frac{1}{3\epsilon}\left(\pi^2+12\left(2\psi(1)-\psi(\tfrac{d}{4})-\psi(\tfrac{d}{2})\right)^2-6\psi_1(\tfrac{d}{2})\right)\right]+\mathcal{O}(\epsilon^0) \,.
\end{split}
}
\end{equation}

The beta function coefficient is
$ \alpha_{T} =\epsilon (4\pi)^{3d/2}\Gamma(\tfrac{d}{2})^3 \frac{(3T-2DS)}{4} $, for which we obtain:
\begin{equation}
\alpha_{T}    = \,  \frac{1}{2}\Big[2\psi(\tfrac{d}{4}) - \psi(\tfrac{d}{2})-\psi(1) \Big]^2 
+\frac{1}{2}\psi_1(1)+  \psi_1(\tfrac{d}{4}) - \frac{3}{2} \psi_1(\tfrac{d}{2}) - \, J_0(\tfrac{d}{4}) \, . 
\label{eq:alpha_5}
\end{equation}

\subsubsection{The $U$ integral}

The $U$ integral is more complicated. We use:
\begin{align}
&\frac{1}
 { [(a_1+a_2)(b_1+b_2)(c_1 + c_2)
+ c_1c_2(a_1+a_2) + a_1a_2 (c_1+c_2) ]^{d/2} }  \crcr
& \qquad = \int_{0^-}[dz_1] \int_{0^-}[dz_2] 
 \frac{\Gamma(-z_1) \Gamma(-z_2) \Gamma(\tfrac{d}{2}+z_1+z_2)}{\Gamma( \tfrac{d}{2})} 
 \frac{ [c_1c_2(a_1+a_2)]^{z_1} [a_1a_2(c_1+c_2)]^{z_2} }
{ [(a_1+a_2)(b_1+b_2)(c_1 + c_2)]^{\tfrac{d}{2} + z_1+z_2} } \,.
\end{align}
Integrating out $a,b,c$ we end up with:
\begin{align}
 \tilde{U}= 
 \frac{1}{\Gamma(\zeta)^4\Gamma(2\zeta)\Gamma(\tfrac{d}{2})} &
 \int_{0^-}[dz_1] \int_{0^-}[dz_2]  \; 
 \Gamma(\tfrac{d}{2}+z_1+z_2) 
 \frac{\Gamma(\zeta + z_1)^2\Gamma(\zeta + z_2)^2}{\Gamma(2\zeta + 2z_1)\Gamma(2\zeta + 2z_2) } \\
 & \times \Gamma(-z_1) \Gamma(-z_2) 
 \Gamma(\tfrac{\epsilon}{2} -z_1-z_2) \Gamma(\tfrac{\epsilon}{2} +z_1 )
 \Gamma( \tfrac{\epsilon}{2}+z_2) \,. \nonumber
\end{align}
The problem now is that, due to the $\Gamma(\epsilon/2 - z_1-z_2)$ factor, the poles in $z_2$ in the right half complex plane depend on $z_1$. This makes the integral quite tricky. The first poles in $z_2$ are located at 
$0$ and $\epsilon/2-z_1$, hence:
\begin{align}
\tilde{U}= &  \frac{1}{\Gamma(\zeta)^4\Gamma(2\zeta)\Gamma( \tfrac{d}{2})} 
 \int_{0^-}[dz_1] \bigg(
 \Gamma( \tfrac{d}{2}+z_1) 
 \frac{\Gamma(\zeta + z_1)^2\Gamma(\zeta )^2}{\Gamma(2\zeta + 2z_1)\Gamma(2\zeta ) } 
 \Gamma(-z_1) \Gamma( \tfrac{\epsilon}{2} -z_1) 
 \Gamma( \tfrac{\epsilon}{2} +z_1 ) \Gamma(\tfrac{\epsilon}{2}) \crcr 
& \qquad    + 
    \Gamma( \tfrac{d}{2}+ \tfrac{\epsilon}{2}) 
 \frac{\Gamma(\zeta + z_1)^2\Gamma(\zeta + \tfrac{\epsilon}{2}-z_1)^2}{\Gamma(2\zeta + 2z_1)\Gamma(2\zeta + \epsilon-2z_1) } \Gamma(-z_1) 
 \Gamma( - \tfrac{\epsilon}{2} +z_1) 
 \Gamma( \tfrac{\epsilon}{2} +z_1 )\Gamma(\epsilon-z_1)  \bigg)
  \crcr
& +  \frac{1}{\Gamma(\zeta)^4\Gamma(2\zeta)\Gamma(\tfrac{d}{2})} 
 \int_{0^-}[dz_1]\int_{1^-}[dz_2]  \; 
 \Gamma( \tfrac{d}{2}+z_1+z_2) 
 \frac{\Gamma(\zeta + z_1)^2\Gamma(\zeta + z_2)^2}{\Gamma(2\zeta + 2z_1)\Gamma(2\zeta + 2z_2) }\crcr
 &  \qquad \qquad \times \Gamma(-z_1) \Gamma(-z_2) 
 \Gamma(\tfrac{\epsilon}{2} -z_1-z_2) \Gamma(\tfrac{\epsilon}{2} +z_1 )
 \Gamma(\tfrac{\epsilon}{2}+z_2) \,. 
\end{align}
We aim to compute this up to order $1/\epsilon$:
\begin{itemize}
 \item the first term is $\tilde D \tilde S$.
\item the second term  has poles at $z_1=0,z_1=\epsilon$ and $z_1=\zeta+\epsilon/2$. The residue at the last pole gives a convergent contribution, thus the divergent part is at most:
\begin{align}
&\frac{\Gamma(\tfrac{d}{2}+ \tfrac{\epsilon}{2})}{\Gamma(\zeta)^4\Gamma(2\zeta)\Gamma( \tfrac{d}{2})} \left [ \frac{\Gamma(\zeta+ \tfrac{\epsilon}{2})^2\Gamma(\zeta)^2\Gamma( \tfrac{\epsilon}{2})\Gamma(\epsilon)\Gamma(-\tfrac{\epsilon}{2})}{\Gamma(2\zeta+\epsilon)\Gamma(2\zeta)} +\frac{\Gamma(\zeta- \tfrac{\epsilon}{2})^2\Gamma(\zeta+\epsilon)^2\Gamma( \tfrac{3\epsilon}{2})\Gamma(-\epsilon)\Gamma(\tfrac{\epsilon}{2})}{\Gamma(2\zeta-\epsilon)\Gamma(2\zeta+2\epsilon)} \right. \nonumber\\
& \qquad \qquad \qquad \qquad \quad\left. - \frac{\Gamma(\zeta)^2\Gamma(\zeta+ \tfrac{\epsilon}{2})^2\Gamma(\tfrac{d}{2}+\tfrac{\epsilon}{2})\Gamma(\epsilon)\Gamma( \tfrac{\epsilon}{2})\Gamma(-\tfrac{\epsilon}{2})}{\Gamma(2\zeta)\Gamma(2\zeta+\epsilon)}\right] \\ 
& + \frac{\Gamma(\tfrac{d}{2}+ \tfrac{\epsilon}{2})}{\Gamma(\zeta)^4\Gamma(2\zeta)\Gamma( \tfrac{d}{2})}\int_{1^-}[dz_1] \; 
 \frac{\Gamma(\zeta + z_1)^2\Gamma(\zeta + \tfrac{\epsilon}{2}-z_1)^2}{\Gamma(2\zeta + 2z_1)\Gamma(2\zeta + \epsilon-2z_1) } \Gamma(-z_1) \Gamma( - \tfrac{\epsilon}{2} +z_1) 
 \Gamma( \tfrac{\epsilon}{2} +z_1 )\Gamma(\epsilon-z_1)  \,, \nonumber
\end{align}
but, using the remark in appendix~\ref{sec:MB} the last integral has a finite limit for $\epsilon \to 0$.
\item the double integral is also tractable. First, we reduce it to
\begin{align}
&  \tilde{D} \frac{ J_{\epsilon}(\zeta) }{\Gamma(\tfrac{d}{2})^2} -   \frac{1}{\Gamma(\zeta)^4\Gamma(2\zeta)\Gamma(\tfrac{d}{2})} 
 \int_{1^-}[dz_2]  \; 
 \Gamma( \tfrac{d}{2} +1+ \tfrac{\epsilon}{2} ) 
 \frac{\Gamma(\zeta + 1+ \tfrac{\epsilon}{2} -z_2 )^2\Gamma(\zeta + z_2)^2}{\Gamma(2\zeta + 2 +\epsilon -2z_2)\Gamma(2\zeta + 2z_2) }\crcr
 &\qquad  \qquad \times \Gamma(-1 - \tfrac{\epsilon}{2}+z_2) \Gamma(-z_2) 
  \Gamma(\epsilon +1-z_2 ) \Gamma(\tfrac{\epsilon}{2}+z_2)  \crcr
  & +  \frac{1}{\Gamma(\zeta)^4\Gamma(2\zeta)\Gamma(\tfrac{d}{2} ) } 
 \int_{1^-}[dz_1]\int_{1^-}[dz_2]  \; 
 \Gamma( \tfrac{d}{2}+z_1+z_2) 
 \frac{\Gamma(\zeta + z_1)^2\Gamma(\zeta + z_2)^2}{\Gamma(2\zeta + 2z_1)\Gamma(2\zeta + 2z_2) }\crcr
 &  \qquad \qquad \times \Gamma(-z_1) \Gamma(-z_2) 
 \Gamma( \tfrac{\epsilon}{2} -z_1-z_2) \Gamma(\tfrac{\epsilon}{2} +z_1 )
 \Gamma( \tfrac{\epsilon}{2}+z_2) \,. 
\end{align}
Again the single integral is $O(\epsilon^0)$.
A detailed study of the remaining double integral shows that only the poles at $z_2=\epsilon/2-z_1+n$ with $n\geq 1$ and $z_1=m_1$ and $z_1=\epsilon+m_1$ with 
$1\le m_1\le n$ respectively $z_1=n+\epsilon/2-m_2 $ and $0\leq m_2 \leq n-1$ can contribute to the singular part but their summed contribution is in fact $O(\epsilon^0)$.
\end{itemize}

We finally have:
\be
\begin{split}
U \, = \, DS + \frac{\Gamma(\tfrac{d}{2}+\tfrac{\epsilon}{2})\Gamma(-\epsilon)\Gamma(\tfrac{\epsilon}{2})\Gamma(\tfrac{3\epsilon}{2})\Gamma(\zeta+\epsilon)^2\Gamma(\zeta-\tfrac{\epsilon}{2})^2}{(4\pi)^{3d/2}\Gamma(\zeta)^4\Gamma(2\zeta)\Gamma(\tfrac{d}{2})\Gamma(2\zeta+2\epsilon)\Gamma(2\zeta-\epsilon)} 
 \, + \, D \frac{J_{\epsilon}(\zeta)}{(4\pi)^d
 \Gamma(\tfrac{d}{2})^2}+\mathcal{O}(\epsilon^0)
\end{split} \,, 
\label{eq:U2_exp}
\ee

which is at the relevant order:
\begin{equation}
\boxed{
\begin{split}
U=\frac{1}{3(4\pi)^{3d/2}\Gamma(d/2)^3}&\Bigg[\frac{8}{\epsilon^3}+\frac{4}{\epsilon^2}\left(5\psi(1)-4\psi(\tfrac{d}{4})-\psi(\tfrac{d}{2})\right)\crcr
& + \frac{1}{\epsilon}\Bigg(-\frac{7\pi^2}{6}-12\psi_1(\tfrac{d}{4})+19\psi_1(\tfrac{d}{2})+12J_0(\tfrac{d}{4}) +19\psi(1)^2-5\psi(\tfrac{d}{2})^2 \crcr
&  \qquad +32\psi(\tfrac{d}{2})\psi(\tfrac{d}{4})-8\psi(\tfrac{d}{4})^2-2\psi(1)(11\psi(\tfrac{d}{2})+8\psi(\tfrac{d}{4})) \Bigg) \Bigg] +\mathcal{O}(\epsilon^0) \,.
\end{split} 
}
\end{equation}

In the beta function we are interested in the combination $\alpha_{U}=\epsilon(4\pi)^{3d/2}\Gamma(d/2)^3\frac{(D^3-4DS+3U)}{4}$:
\begin{equation}
\alpha_{U}=-\psi_1(1)-\psi_1(\tfrac{d}{4})+2\psi_1(\tfrac{d}{2})+J_0(\tfrac{d}{4}) \, .
\label{eq:alpha_3}
\end{equation}

\subsubsection{The $I_1$ integral}

Let us now compute $I_1$. First we can simplify it by introducing $a_1=a\beta, a_2=a(1-\beta)$ and integrating out $\beta$. We obtain:
\begin{equation}
\tilde{I}_1 = \frac{1}{\Gamma(\zeta)^4\Gamma(2\zeta)} \int_0^{\infty} 
 \frac{ a^{2\zeta-1}( b c_1 c_{2'}c_{2''})^{\zeta-1} e^{-(a+b+c_1+c_{2'}+c_{2''})}}
 { \big[ a(b+ c_1)( c_{2'} +c_{2''} ) + bc_1( c_{2'} + c_{2''}) +ac_{2'}c_{2''}  + b c_{2'}c_{2''}   \big]^{d/2}} \,.
\end{equation}
We split the denominator using three Mellin parameters:
\begin{align}
&\frac{1}{\big[ a(b+ c_1)( c_{2'} +c_{2''} ) + bc_1( c_{2'} + c_{2''}) +ac_{2'}c_{2''}  + b c_{2'}c_{2''}   \big]^{d/2}}
  = \int_{0^-}[dz_1] \int_{0^-}[dz_2] \int_{0^-}[dz_3] 
\crcr
& \qquad
 \frac{\Gamma(-z_1) \Gamma(-z_2)\Gamma(-z_3) 
 \Gamma(\tfrac{ d}{2} +z_1+z_2+z_3)}{\Gamma( \tfrac{d}{2})} \; 
 \frac{ [ac_{2'}c_{2''}]^{z_1} [bc_{2'}c_{2''})]^{z_2}[bc_1(c_{2'}+c_{2''})]^{z_3} }
{ [a(b+c_1)(c_{2'} + c_{2''})]^{d/2 + z_1+z_2+z_3} } \; 
 \,,
\end{align}
and integrating out the Schwinger parameters we find:
\begin{align}
& \tilde{I}_1=\frac{1}{\Gamma(\zeta)^4\Gamma(2\zeta)\Gamma(\tfrac{d}{2})}\int_{0^-}[dz_1] \int_{0^-}[dz_2] \int_{0^-}[dz_3]  \Gamma(-z_1)\Gamma(-z_2)\Gamma(-z_3)\Gamma(\tfrac{d}{2}+z_1+z_2+z_3) \crcr
& \qquad \qquad \times
\frac{\Gamma(\zeta+z_1+z_2)^2\Gamma(\tfrac{\epsilon}{2}+z_1+z_2)}{\Gamma(2\zeta+2z_1+2z_2)}  \frac{\Gamma(\zeta+z_2+z_3)\Gamma(\zeta+z_3)}{\Gamma(2\zeta+2z_3+z_2)} \crcr
&\qquad \qquad \qquad \times\Gamma(\tfrac{\epsilon}{2}-z_2-z_3)\Gamma(\tfrac{\epsilon}{2}+z_3-z_1) \,.
\end{align}
Again, we have a mixing between the poles of the $z_i$. The first poles of $z_3$ are located at $0$ and $\epsilon/2-z_2$ and we have:
\begin{align}
 \tilde{I}_1=& \frac{1}{\Gamma(\zeta)^4\Gamma(2\zeta)\Gamma( \tfrac{ d}{2})}\int_{0^-}[dz_1] \int_{0^-}[dz_2]
\Gamma(-z_1)\Gamma(-z_2) \crcr
& \times \bigg[  \Gamma(\tfrac{ d}{2} +z_1+z_2) \frac{\Gamma(\zeta+z_1+z_2)^2\Gamma(\tfrac{\epsilon}{2}+z_1+z_2)}{\Gamma(2\zeta+2z_1+2z_2)} \; \frac{\Gamma(\zeta+z_2)\Gamma(\zeta)}{\Gamma(2\zeta+z_2)}\Gamma(\tfrac{\epsilon}{2}-z_2)\Gamma(\tfrac{\epsilon}{2}-z_1) \crcr
& \;\;+  \Gamma(z_2-\tfrac{\epsilon}{2})\Gamma(\tfrac{ d}{2} +z_1+\tfrac{\epsilon}{2})  \; \frac{\Gamma(\zeta+z_1+z_2)^2\Gamma(\tfrac{\epsilon}{2}+z_1+z_2)}{\Gamma(2\zeta+2z_1+2z_2)}  \;  \frac{\Gamma(\zeta+\tfrac{\epsilon}{2})\Gamma(\zeta+\tfrac{\epsilon}{2}-z_2)}{\Gamma(2\zeta+\epsilon-z_2)}\Gamma(\epsilon-z_1-z_2) \bigg] \crcr
&+ \frac{1}{\Gamma(\zeta)^4\Gamma(2\zeta)\Gamma(\tfrac{d}{2} )}\int_{0^-}[dz_1] \int_{0^-}[dz_2] \int_{1^-}[dz_3] \Gamma(-z_1)\Gamma(-z_2)\Gamma(-z_3)\Gamma(\tfrac{d}{2}+z_1+z_2+z_3) \crcr
& \qquad \times \frac{\Gamma(\zeta+z_1+z_2)^2\Gamma(\tfrac{\epsilon}{2}+z_1+z_2)}{\Gamma(2\zeta+2z_1+2z_2)}  \; \frac{\Gamma(\zeta+z_2+z_3)\Gamma(\zeta+z_3)}{\Gamma(2\zeta+2z_3+z_2)}\crcr
&\qquad \qquad  \times\Gamma(\tfrac{\epsilon}{2}-z_2-z_3)\Gamma(\tfrac{\epsilon}{2}+z_3-z_1) \,.
\label{eq:I1_double_int}
\end{align}

Let us call $\tilde{I}_{1,1}$ the first double integral. 
The computation of this integral is very similar to the one of $T$. The poles of $z_1$ and $z_2$ are completely independent. Only the poles at $(0,0),(0,\epsilon/2);(\epsilon/2,0),(\epsilon/2,\epsilon/2)$ give singular contributions and we get:
\begin{align}
\tilde{I}_{1,1}=&\frac{\Gamma(\tfrac{\epsilon}{2})^3}{\Gamma(2\zeta)^3}+\frac{\Gamma(\zeta+\tfrac{\epsilon}{2})^3\Gamma(\tfrac{d}{2}+\tfrac{\epsilon}{2})}{\Gamma(\zeta)^3\Gamma(2\zeta)\Gamma(\tfrac{d}{2})\Gamma(2\zeta+\epsilon)\Gamma(2\zeta+\tfrac{\epsilon}{2})}\Gamma(-\tfrac{\epsilon}{2})\Gamma(\epsilon)\Gamma(\tfrac{\epsilon}{2}) \crcr
&+ \frac{\Gamma(\zeta+\tfrac{\epsilon}{2})^2\Gamma(\tfrac{d}{2}+\tfrac{\epsilon}{2})}{\Gamma(\zeta)^2\Gamma( \tfrac{d}{2})\Gamma(2\zeta)^2\Gamma(2\zeta+\epsilon)}\Gamma(-\tfrac{\epsilon}{2})\Gamma(\epsilon)\Gamma(\tfrac{\epsilon}{2}) \crcr
& + \frac{\Gamma(\zeta+\epsilon)^2\Gamma(\zeta+\tfrac{\epsilon}{2})\Gamma(\tfrac{d}{2}+\epsilon)}{\Gamma(\zeta)^3\Gamma(d/2)\Gamma(2\zeta)\Gamma(2\zeta+2\epsilon)\Gamma(2\zeta+\tfrac{\epsilon}{2})}\Gamma(-\tfrac{\epsilon}{2})^2\Gamma(\tfrac{3\epsilon}{2}) +\mathcal{O}(\epsilon^0) \,.
\end{align}
Let us now call $\tilde{I}_{1,2}$ the second term in \eqref{eq:I1_double_int}.
The poles in $z_1$ and $z_2$ mix and the computation is similar to the one of $U$.  The first poles in $z_2$ are located at $z_2=0$, $z_2=\epsilon/2$ and $z_2=\epsilon-z_1$. Pushing the integral over $z_2$ past these poles, we obtain singular contributions only from the single integrals (and in particular from the poles in $z_1$ closest to zero). In the end we obtain:
\begin{align}
 \tilde{I}_{1,2}=\frac{1}{\Gamma(\zeta)^4\Gamma(2\zeta)\Gamma(\tfrac{d}{2})}&\bigg[  \frac{\Gamma(\zeta)^2\Gamma(\zeta+\tfrac{\epsilon}{2})^2\Gamma(\tfrac{d}{2}+\tfrac{\epsilon}{2})}{\Gamma(2\zeta)\Gamma(2\zeta+\epsilon)}\Gamma(-\tfrac{\epsilon}{2})\Gamma(\tfrac{\epsilon}{2})\Gamma(\epsilon) \crcr
& - \frac{\Gamma(\zeta)\Gamma(\zeta+\tfrac{\epsilon}{2})^3\Gamma(\tfrac{d}{2}+\tfrac{\epsilon}{2})}{\Gamma(2\zeta+\tfrac{\epsilon}{2})\Gamma(2\zeta+\epsilon)}\Gamma(-\tfrac{\epsilon}{2})\Gamma(\tfrac{\epsilon}{2})\Gamma(\epsilon) \crcr
&+ \frac{\Gamma(\zeta+\epsilon)^2\Gamma(\zeta+\tfrac{\epsilon}{2})\Gamma(\zeta-\tfrac{\epsilon}{2})\Gamma(\tfrac{d}{2}+\tfrac{\epsilon}{2})}{\Gamma(2\zeta)\Gamma(2\zeta+2\epsilon)}\Gamma(\tfrac{\epsilon}{2})\Gamma(\tfrac{3\epsilon}{2})\Gamma(-\epsilon)   \bigg] \crcr
& + \frac{\Gamma(\tfrac{3\epsilon}{2})}{\Gamma(\tfrac{d}{4})\Gamma(\tfrac{d}{2})^3}\int_{1^-} [dz_1] \Gamma(-z_1)^2\Gamma(z_1)\Gamma(\tfrac{d}{4}+z_1)+\mathcal{O}(\epsilon^0)\, .
\end{align}
The triple integral in \eqref{eq:I1_double_int}, which we call $\tilde{I}_{1,3}$ has poles in $z_2$ between $0^-$ and $1^-$ located at $0$ and $1+\epsilon/2-z_3$. Pushing the contour of  $z_2$ past these poles, a very lengthy but straightforward computing gives:
\begin{equation}
\tilde{I}_{1,3}= \tilde{D} \;\frac{ J_{\epsilon}(\zeta) }{\Gamma(\tfrac{d}{2} )^2}+ \mathcal{O}(\epsilon^0)\, .
\end{equation}
Gathering all the terms from $\tilde{I}_{1,1}$, $\tilde{I}_{1,2}$ and $\tilde{I}_{1,3}$, we have:
\begin{equation}
\begin{split}
I_1=&\frac{\Gamma(\tfrac{\epsilon}{2})^3 }{(4\pi)^{\tfrac{3d}{2}}\Gamma(2\zeta)^3} +\frac{\Gamma(\zeta+\epsilon)^2\Gamma(\zeta+\tfrac{\epsilon}{2})\Gamma(\zeta-\tfrac{\epsilon}{2})\Gamma(\tfrac{d}{2}+\tfrac{\epsilon}{2})}{(4\pi)^{\tfrac{3d}{2}}\Gamma(\zeta)^4\Gamma(2\zeta)^2\Gamma(\tfrac{d}{2})\Gamma(2\zeta+2\epsilon)}\Gamma(\tfrac{\epsilon}{2})\Gamma(\tfrac{3\epsilon}{2})\Gamma(-\epsilon) \crcr
&+ \frac{\Gamma(\zeta+\epsilon)^2\Gamma(\zeta+\tfrac{\epsilon}{2})\Gamma(\tfrac{d}{2}+\epsilon)}{(4\pi)^{\tfrac{3d}{2}}\Gamma(\zeta)^3\Gamma(\tfrac{d}{2})\Gamma(2\zeta)\Gamma(2\zeta+2\epsilon)\Gamma(2\zeta+\tfrac{\epsilon}{2})}\Gamma(-\tfrac{\epsilon}{2})^2\Gamma(\tfrac{3\epsilon}{2}) \crcr
& +\frac{2\Gamma(\zeta+\tfrac{\epsilon}{2})^2\Gamma(\tfrac{d}{2}+\tfrac{\epsilon}{2})}{(4\pi)^{\tfrac{3d}{2}}\Gamma(\zeta)^2\Gamma(\tfrac{d}{2})\Gamma(2\zeta)^2\Gamma(2\zeta+\epsilon)}\Gamma(-\tfrac{\epsilon}{2})\Gamma(\epsilon)\Gamma(\tfrac{\epsilon}{2})    \crcr
& + \Gamma(\tfrac{3\epsilon}{2})\frac{1}{(4\pi)^{\tfrac{3d}{2}}\Gamma(\tfrac{d}{4})\Gamma(\tfrac{d}{2})^3}\int_{1^-} [dz_1] \Gamma(-z_1)^2\Gamma(z_1)\Gamma(\tfrac{d}{4}+z_1) + D \; \frac{ J_{\epsilon}(\zeta) }{\Gamma(\tfrac{d}{2})^2}+\mathcal{O}(\epsilon^0) \,.
\end{split}
\end{equation}
The last step is to compute the finite integral in this equation. We close the contour to the left and pick up a pole of order $3$ at $z=0$ and poles of order one at $z=-n_1$ and $z=-d/4-n_2$ with $n_1\geq 1$ and $n_2\geq 0$. The sums over $n_1$ and $n_2$ can be computed in terms of gamma and polygamma functions and we find:
\begin{equation}
\begin{split}
& \frac{1}{\Gamma( \tfrac{d}{4} )\Gamma(\tfrac{d}{2})^3}\int_{1^-} [dz] \Gamma(-z)^2\Gamma(z)\Gamma(\tfrac{d}{4}+z) \crcr
&\;\;  = \frac{1}{\Gamma(\tfrac{d}{2})^3}\left(-\pi^2\csc(\tfrac{d\pi}{4})^2+\psi_1(1-\tfrac{d}{4})+\frac{1}{2}\psi_1(\tfrac{d}{4})+\frac{1}{2}\psi(1)^2+\frac{\pi^2}{4}-\psi(1)\psi(\tfrac{d}{4})+\frac{1}{2}\psi(\tfrac{d}{4})^2\right) \,.
\end{split}
\end{equation}

We finally have for $I_1$:

\begin{equation}
\boxed{
\begin{split}
I_1=\frac{1}{(4\pi)^{3d/2}\Gamma(d/2)^3}\Bigg[&\frac{4}{3\epsilon^3}-\frac{4}{\epsilon^2}\left(\psi(\tfrac{d}{4})-\psi(1)\right) \crcr
& +\frac{1}{\epsilon}\Bigg(2J_0(\tfrac{d}{4})-\frac{\pi^2}{9}+5\psi(1)^2-\psi(\tfrac{d}{2})^2+4\psi(\tfrac{d}{4})\psi(\tfrac{d}{2})+2\psi(\tfrac{d}{4})^2\crcr
&  -2\psi(1)(\psi(\tfrac{d}{2})+4\psi(\tfrac{d}{4}))-2\psi_1(\tfrac{d}{4})+\frac{8}{3}\psi_1(\tfrac{d}{2}) \Bigg) \Bigg] +\mathcal{O}(\epsilon^0) \,,
\end{split}
}
\end{equation}

and finally we get for $\alpha_{I_1}$:
\begin{align}
\alpha_{I_1}=\frac{3}{2}\left[2\psi(\tfrac{d}{4})-\psi(\tfrac{d}{2})-\psi(1)\right]^2+\frac{1}{2}\psi_1(1)-\frac{1}{2}\psi_1(\tfrac{d}{2}) \, .
\end{align}

\subsubsection{The $I_3$ integral}

For the $I_3$ diagram, one needs to take into account the subtraction of the local part of the two-point insertion. After some trivial integrals, the subtracted $I_3$ writes 
using a Taylor expansion with integral rest:
\begin{equation}
I_3=\frac{ \tfrac{-2}{d} }{(4\pi)^{3d/2}\Gamma(3\zeta)\Gamma(\zeta)^3}\int_0^1 dt\int [dadb] \frac{a^{3\zeta-1}(b_1b_2b_3)^{\zeta}e^{-a-b_1-b_2-b_3}}{\left[a(b_1b_2+b_1b_3+b_2b_3)+tb_1b_2b_3\right]^{d/2+1}} \,.
\end{equation}
We then introduce Mellin parameters via
\begin{align}
&\frac{1}{\left[a(b_2b_3+b_1b_2+b_1b_3)+tb_1b_2b_3\right]^{1+d/2}}\crcr
& =\int_{0^-}[dz_1] \int_{0^-} [dz_2]\frac{\Gamma(-z_1)\Gamma(-z_2)\Gamma(z_1+z_2+d/2+1)}{\Gamma(1+d/2)}\frac{(tb_1b_2b_3)^{z_1}(a(b_2b_3))^{z_2}}{\left[ab_1(b_2+b_3)\right]^{z_1+z_2+d/2+1}} \,,
\end{align}
and integrate out the Schwinger parameters and $t$  to obtain:
\begin{align}
I_3=&\frac{ \tfrac{-2}{d} }{(4\pi)^{3d/2}\Gamma(3\zeta)\Gamma(\zeta)^3\Gamma(1+\tfrac{d}{2})}\int_{(\tfrac{d+3\epsilon}{4}-1)^-} [dz_1] \int_{(\tfrac{-d+\epsilon}{4})^-} [dz_2]  \Gamma(-z_1)\Gamma(-z_2)\Gamma(3\zeta-\tfrac{d}{2}-1-z_1)\crcr 
&\Gamma(\zeta-\tfrac{d}{2}-z_2) \Gamma(1+\tfrac{d}{2}+z_1+z_2)\frac{\Gamma(\zeta+1+z_1+z_2)^2}{\Gamma(2\zeta+2+2z_1+2z_2)(z_1+1)}\Gamma(\tfrac{\epsilon}{2}+1+z_1+z_2) \,,
\end{align}
where we moved the contours such that all Gamma functions have positive arguments (i.e. the integrals over the Schwinger parameters are convergent) and the poles of the first four Gamma functions are separated from the ones of the other Gamma functions by the contours. The pole in $z_1$ and $z_2$ are independent, and the only one contributing at order ${\cal O}(\epsilon^{-1})$ is located at $z_1=3\zeta-d/2-1$ and $z_2=\zeta-d/2$.  We finally obtain:
\begin{equation}
\boxed{ I_3=\frac{2}{3\epsilon}\frac{\Gamma(- \tfrac{d}{4})}{(4\pi)^{3d/2}\Gamma( \tfrac{3d}{4} )\Gamma( \tfrac{d}{2})}+\mathcal{O}(\epsilon^0)\,, }
\end{equation}
and for $\alpha_{I_3}= \epsilon (4\pi)^{3d/2}\Gamma(\tfrac{d}{2})^3 \frac{I_3}{2}$ we get:
\begin{equation}
\alpha_{I_3}=\frac{\Gamma(- \tfrac{d}{4})
\Gamma( \tfrac{d}{2} )^2}{3 \, \Gamma( \tfrac{3d}{4} )} \,. 
\end{equation}

\subsubsection{The $I_4$ integral}
\label{app:I_4}

We will compute the $I_4$ integral differently.
The Feynman integral corresponding to the tetrahedron diagram at zero external momenta is:
\begin{equation}
\begin{split}
& \mu^{-3\epsilon}I_4= \int \, \frac{d^d q_1}{(2\pi)^d} \frac{d^d q_2}{(2\pi)^d}  \frac{d^d q_3}{(2\pi)^d}  \crcr 
& \frac{1}{(q_1^2+\mu^2)^{\zeta}(q_2^2+\mu^2)^{\zeta}(q_3+\mu^2)^{\zeta}((q_1-q_2)^2+\mu^2)^{\zeta}((q_3-q_1)^2+\mu^2)^{\zeta}((q_3-q_2)^2+\mu^2)^{\zeta}} \,.
\end{split}
\end{equation}
This is a much harder diagram to compute with the Schwinger parametrization and Mellin-Barnes representation, so we will adopt a different approach, following the method discussed in \cite{Kreimer:1996js} for the case $\zeta=1$, and based on the Gegenbauer polynomial technique \cite{Chetyrkin:1980pr}.
As this diagram does not contain divergent subgraphs, it diverges as a simple pole in $\epsilon$ and we are only interested in determining the residue at the pole. The ultraviolet divergence arises when all three loop momenta are large, and its coefficient is independent of the chosen IR regularization: we can set $\mu=0$ and deal with the infrared divergence in a simpler fashion.
The main trick is to use the following expansion, valid  for $p>q$:
\be
\frac{1}{(p-q)^{2\zeta}} = \frac{1}{p^{2\zeta}} \sum_{n= 0}^{+\infty} C_n^{\zeta}(\hat{p}\cdot\hat{q}) \left(\frac{q}{p}\right)^n \,,
\ee
where $p=\sqrt{p^2}$ and $\hat{p}_\mu = p_\mu/p$. The expansion coefficients $C_n^\zeta(x)$ are the Gegenbauer polynomials, satisfying
\be
C_n^{\zeta}(1)=  \frac{\Gamma(n+2\zeta)}{\Gamma(2\zeta) n!} \,, \qquad
\int d\hat{q}\; C_n^{\zeta}(\hat{p}\cdot\hat{q}) C_{n'}^{\zeta}(\hat{p}'\cdot\hat{q}) = \frac{\zeta}{n+\zeta} \delta_{nn'} C_n^{\zeta}(\hat{p}\cdot\hat{p}') \,,
\ee
where the angular integral is normalized such that $\int d\hat{q} =1$.

Noticing that $I_4$ is totally symmetric in the three loop momenta, we can choose $q_1<q_2<q_3$ and write:
\begin{equation}
\begin{split}
I_4 &=6\mu^{3\epsilon} \int_{q_1<q_2<q_3} \, \frac{d^d q_1}{(2\pi)^d} \frac{d^d q_2}{(2\pi)^d}  \frac{d^d q_3}{(2\pi)^d}  \frac{1}{q_1^{2\zeta}q_2^{2\zeta}q_3^{2\zeta}(q_1-q_2)^{2\zeta}(q_3-q_1)^{2\zeta}(q_3-q_2)^{2\zeta}} \\
&= 6 \mu^{3\epsilon} \int_{q_1<q_2<q_3} \, \frac{d^d q_1}{(2\pi)^d} \frac{d^d q_2}{(2\pi)^d}   \frac{d^d q_3}{(2\pi)^d}   \frac{1}{q_1^{2\zeta}q_2^{4\zeta}q_3^{6\zeta}}\\
&\quad\qquad \times \sum_{n_1,n_2,n_3} C_{n_1}^{\zeta}(\hat{q_1}\cdot\hat{q_2}) \left(\frac{q_1}{q_2}\right)^{n_1} C_{n_2}^{\zeta}(\hat{q_1}\cdot\hat{q_3}) \left(\frac{q_1}{q_3}\right)^{n_2} C_{n_3}^{\zeta}(\hat{q_3}\cdot\hat{q_2}) \left(\frac{q_2}{q_3}\right)^{n_3} \,.
\end{split}
\end{equation}
Separating into radial and angular integrals, this becomes:
\begin{equation}
\begin{split}
I_4=& \frac{6\mu^{3\epsilon}  {\rm Vol}(S^{d-1})^3}{(2\pi)^{3d}} \int_{q_1<q_2<q_3} \, d q_1d q_2 d q_3  \, q_1^{d-1-2\zeta}q_2^{d-1-4\zeta}q_3^{d-1-6\zeta}\\
&\quad\qquad \times \sum_{n_1,n_2,n_3} \int d \hat{q}_1d \hat{q}_2 d \hat{q}_3 \, C_{n_1}^{\zeta}(\hat{q_1}\cdot\hat{q_2}) \left(\frac{q_1}{q_2}\right)^{n_1} C_{n_2}^{\zeta}(\hat{q_1}\cdot\hat{q_3}) \left(\frac{q_1}{q_3}\right)^{n_2} C_{n_3}^{\zeta}(\hat{q_3}\cdot\hat{q_2}) \left(\frac{q_2}{q_3}\right)^{n_3} \,,
\end{split}
\end{equation}
and using the orthogonality relation of the Gegenbauer polynomials we get:
\begin{equation}
\begin{split}
I_4 
&= \frac{6 \mu^{3\epsilon} {\rm Vol}(S^{d-1})^3}{(2\pi)^{3d}} \int_{q_1<q_2<q_3} \, d q_1d q_2 d q_3  \, q_1^{d-1-2\zeta}q_2^{d-1-4\zeta}q_3^{d-1-6\zeta}\\
&\quad\qquad \times \sum_{n_1,n_2}  \frac{\zeta}{n_1+\zeta} \int d \hat{q}_1  d \hat{q}_3 \, C_{n_1}^{\zeta}(\hat{q_1}\cdot\hat{q_3}) C_{n_2}^{\zeta}(\hat{q_1}\cdot\hat{q_3}) \left(\frac{q_1}{q_3}\right)^{n_1+n_2} \\
&= \frac{6 \mu^{3\epsilon} {\rm Vol}(S^{d-1})^3}{(2\pi)^{3d}} \int_{\mu}^{+\infty} d q_3 \int_{\mu}^{q_3} d q_2  \int_{\mu}^{q_2}  d q_1  \, q_1^{d-1-2\zeta}q_2^{d-1-4\zeta}q_3^{d-1-6\zeta} \crcr& \qquad \qquad \times \sum_{n\geq 0}  \left(\frac{\zeta}{n+\zeta}\right)^2 \frac{\Gamma(n+2\zeta)}{\Gamma(2\zeta) n!}  \left(\frac{q_1}{q_3}\right)^{2 n} \\
&= \frac{6  {\rm Vol}(S^{d-1})^3}{(2\pi)^{3d}} \frac{1}{\epsilon} \frac{ (\tfrac{d}{4})^2}{12 
\Gamma(\tfrac{d}{2})} \sum_{n\geq 0}  \frac{1}{(n+
\tfrac{d}{4} )^4} \frac{\Gamma(n+ \tfrac{d}{2})}{n!}  + O(\epsilon^0) \,.
\end{split}
\end{equation}
We finally obtain $I_4$ remembering that ${\rm Vol}(S^{d-1})= 2\pi^{d/2}/\Gamma(\tfrac{d}{2})$,  and noting that 
the last sum can be written in terms of polygamma functions:
\be
\boxed{ I_4 = \frac{1}{\epsilon} \frac{d^2 }{48 (4\pi)^{3d/2}}   
\frac{\Gamma(\tfrac{d}{4})^3\Gamma(1-\tfrac{d}{4})}{\Gamma(\tfrac{d}{2})^4}(6\psi_1(\tfrac{d}{4})-\pi^2) + O(\epsilon^0) \,. }
\ee
Notice that for $d=4$ this reduces to $I_4 = \frac{{\rm Vol}(S^{d-1})^3}{2 (2\pi)^{3d} \epsilon} \zeta(3)$, in agreement with \cite{Brezin:1974-add}.

In the beta functions we are interested in $\alpha_{I_4}=\epsilon(4\pi)^{3d/2}\Gamma(\tfrac{d}{2})^3 3I_4$. We obtain:

%

\begin{equation}
\alpha_{I_4} \, = \,\frac{\, \Gamma(1 + \tfrac{d}{4})^3\Gamma(- \tfrac{d}{4})}{
 \, \Gamma(\tfrac{d}{2} )} \; 6 \, \Big[  \psi_1(1) -  \psi_1(\tfrac{d}{4})  \Big] \, .
\end{equation}

\addcontentsline{toc}{section}{References}

\bibliographystyle{JHEP}
\bibliography{Refs-TMV,Refs-CFT,Refs-QFT}


\end{document}